\documentclass[3p,times,twocolumn]{elsarticle}

\usepackage{ecrc}

\volume{00}

\firstpage{1}

\journalname{Nuclear Inst. and Methods in Physics Research, A}

\runauth{J. Ripa et al.}

\jid{nima}

\jnltitlelogo{NIMA}

\usepackage{amssymb}

\usepackage[hyphens]{url}

\biboptions{sort&compress}

\usepackage[figuresright]{rotating}

\begin{document}

\begin{frontmatter}


\cortext[cor1]{Email: ripa.jakub@gmail.com}


\dochead{}

\title{
\hfill \break
\hfill \break
Characterization of more than three years of in-orbit radiation damage of SiPMs on GRBAlpha and VZLUSAT-2 CubeSats}


\author[mu]{Jakub Řípa\corref{cor1}}
\author[mu]{Marianna Dafčíková}
\author[mu]{Pavel Kosík}
\author[mu]{Filip Münz}
\author[hi]{Masanori Ohno}
\author[wi]{Gábor Galgóczi}
\author[mu]{Norbert Werner}
\author[ko]{András Pál}
\author[ko]{László Mészáros}
\author[ko]{Balázs Csák}

\author[hi]{Yasushi Fukazawa}
\author[hi]{Hiromitsu Takahashi}
\author[hi]{Tsunefumi Mizuno}
\author[na]{Kazuhiro Nakazawa}
\author[ut]{Hirokazu Odaka}
\author[ri]{Yuto Ichinohe}

\author[sp]{Jakub Kapuš}
\author[sp]{Jan Hudec}
\author[sp]{Marcel Frajt}
\author[sp]{Maksim Rezenov}

\author[vz]{Vladimír Dániel}
\author[vz]{Petr Svoboda}
\author[vz]{Juraj Dudáš}
\author[vz]{Martin Sabol}

\author[ne]{Róbert László}
\author[ne]{Martin Koleda}

\author[mu]{Michaela Ďuríšková}
\author[mu]{Lea Szakszonová}
\author[mu]{Martin Kolář}
\author[mu]{Nikola Husáriková}
\author[hi,mu]{Jean-Paul Breuer}
\author[mu]{Filip Hroch}
\author[mu]{Tomáš Vítek}

\author[zcu]{Ivo Veřtát}
\author[vut]{Tomáš Urbanec}
\author[vut]{Aleš Povalač}
\author[vut]{Miroslav Kasal}

\author[tuk]{Peter Hanák}
\author[edis]{Miroslav Šmelko}

\author[ca]{Martin Topinka}

\author[nthu-astr]{Hsiang-Kuang Chang}
\author[nycu]{Tsung-Che Liu}
\author[as]{Chih-Hsun Lin}
\author[ncue]{Chin-Ping Hu}
\author[nthu-pme]{Che-Chih Tsao}

\address[mu]{Department of Theoretical Physics and Astrophysics, Faculty of Science, Masaryk University, Czech Republic}
\address[wi]{Wigner Research Centre for Physics, Budapest, Hungary}
\address[ko]{Konkoly Observatory, Research Centre for Astronomy and Earth Sciences, Budapest, Hungary}
\address[hi]{Department of Physics, Graduate School of Advanced Science and Engineering, Hiroshima University, Higashi-Hiroshima, Japan}
\address[na]{Department of Physics, Nagoya University, Nagoya, Aichi, Japan}
\address[vz]{Czech Aerospace Research Centre, Prague, Czech Republic}
\address[sp]{Spacemanic Ltd., Bratislava, Slovakia}
\address[ut]{Department of Physics, The University of Tokyo, Bunkyo-ku, Tokyo, Japan}
\address[ri]{RIKEN Nishina Center for Accelerator-Based Science, Saitama, Japan}
\address[zcu]{University of West Bohemia, Department of Applied Electronics and Telecommunications, Plzeň, Czech Republic}
\address[vut]{Department of Radio Electronics, Faculty of Electrical Engineering and Communication, Brno University of Technology, Brno, Czech Republic}
\address[tuk]{Faculty of Aeronautics, Technical University of Kosice, Košice, Slovakia}
\address[ne]{Needronix Ltd., Bratislava, Slovakia}
\address[ca]{INAF - Osservatorio Astronomico di Cagliari, Selargius (CA), Italy}
\address[nthu-astr]{Institute of Astronomy, National Tsing Hua University, Hsinchu, Taiwan, Republic of China}
\address[nycu]{Institute of Physics, National Yang Ming Chiao Tung University, Hsinchu, Taiwan, Republic of China}
\address[as]{Institute of Physics, Academia Sinica, Taipei, Taiwan, Republic of China}
\address[ncue]{Department of Physics, National Changhua University of Education, Changhua City, Taiwan, Republic of China}
\address[nthu-pme]{Department of Power Mechanical Engineering, National Tsing Hua University, Hsinchu, Taiwan, Republic of China}
\address[edis]{EDIS vvd., Košice, Slovakia}

\begin{abstract}
Silicon photomultipliers (SiPMs) are prone to radiation damage which causes an increase of dark count rate. This leads to an increase in low-energy threshold in a gamma-ray detector combining SiPM and a scintillator. Despite this drawback, they are becoming preferred for scintillator-based gamma-ray detectors on CubeSats due to their low operation voltage, small size, linear response to low light intensity and fast response. This increasing popularity of SiPMs among new spaceborne missions makes it important to characterize their long-term performance in the space environment. In this work, we report the change of the dark count rate and low-energy threshold of S13360-3050 PE multi-pixel photon counters (MPPCs) by Hamamatsu, using measurements acquired by the GRBAlpha and VZLUSAT-2 CubeSats at low Earth orbit (LEO) spanning over three years. Such a long measurement of the performance of MPPCs in space has not been published before. GRBAlpha is a 1U CubeSat launched on March 22, 2021, to a 550 km altitude sun-synchronous polar orbit (SSO) carrying on board a gamma-ray detector based on CsI(Tl) scintillator readout by eight MPPCs and regularly detecting gamma-ray transients such as gamma-ray bursts and solar flares in the energy range of $\sim30-900$\,keV. VZLUSAT-2 is a 3U CubeSat launched on January 13, 2022 also to a 535\,km altitude SSO carrying on board, among other payloads, two gamma-ray detectors similar to the one on GRBAlpha. We have flight-proven the Hamamatsu MPPCs S13360-3050 PE and demonstrated that MPPCs, shielded by 2.5\,mm of PbSb alloy, can be used in LEO environment on a scientific mission lasting beyond three years. This manifests the potential of MPPCs being employed in future satellites.
\end{abstract}

\begin{keyword}
SiPM \sep MPPC \sep Radiation damage \sep Space environment \sep LEO \sep CubeSat
\end{keyword}

\end{frontmatter}



\section{Introduction}
\label{sec:intro}
Silicon photomultipliers (SiPMs) are arrays of avalanche photodiodes (APDs) operating in Geiger mode which are capable of registering individual photons \citep{2019NIMPA.926...16A, 2019NIMPA.926....2P, 2019NIMPA.926...36K, 2011NIMPA.656...69V}. Their low operation voltage, small size, linear response to the input low light intensity and fast response comparable to classical photomultiplier tubes make them favourable for scintillator-based gamma-ray detectors, particularly on CubeSats and small spaceborne instruments with limited volume and power. However, it is well known that SiPMs are prone to radiation damage. For example work \citep{2003NIMPA.512...30L} describes the radiation damage effects in silicon detectors irradiated by hadron and gamma radiation and discusses the impact on detector properties and microscopic defects. For a review of the radiation tolerance of SiPMs see work \citep{2019NIMPA.926...69G}. Several experiments has shown that radiation-damaged SiPMs have deteriorated performance \citep{2023ITNS...70..150S, 2023NIMPA104567488A, 2023NIMPA104767791A, 2023ExA....55..343M, 2021NIMPA.98864798M, 2019ITNS...66.1833B}.

With the increasing popularity of SiPMs among new spaceborne missions it is of paramount importance to characterize their performance in space, e.g. the evolution of their dark count rate or sensitivity energy range of a scintillator-based detector which employs SiPMs.
A number of new X-ray/gamma-ray missions for high-energy astrophysics which utilize SiPMs in their detectors are being developed, have recently been launched or are planned. Therefore, in-orbit measurements of the radiation-induced ageing of SiPMs over several years is highly valuable. Current and future CubeSats, larger missions and instruments using SiPMs for detection of gamma-ray transients, such as gamma-ray bursts (GRBs) \citep{2006RPPh...69.2259M} are for example \citep{2022hxga.book...19B, 2021Galax...9..120F}:
BurstCube \citep{2017ICRC...35..760P}, EIRSAT-1 \citep{2021ExA....52...59M,2022ExA....53..961M}, GALI \citep{10.1117/12.3019170}, GARI \citep{2023SPIE12678E..15M}, GECAM/GRD \citep{2023NIMPA105668586Z}, GlowBug \citep{2022SPIE12181E..1OW}, GRBAlpha \citep{2023A&A...677A..40P}, GRBBeta \citep{10.1117/12.3025855}, GRID \citep{2019ExA....48...77W}, GTM \citep{2022AdSpR..69.1249C}, MAMBO \citep{2022SPIE12181E..2LB}, POLAR-2 \citep{2025arXiv250107758K}, SIRI \citep{2019SPIE11118E..0IM}, StarBurst \citep{2024NIMPA106469329W} or VZLUSAT-2 \citep{2020SPIE11530E..0ZD}.

The SIRI-1 instrument on board the STPSat-5 on low Earth orbit (LEO) reported a significant and constant increase of leakage current over one year of J-series SensL 60035 SiPMs used in their detector \citep{2021NIMPA.98864798M}. GRID-02 CubeSat characterized the ageing of ON Semiconductor MicroFJ-60035-TSV SiPMs in orbit over three months and also reported a significant and gradual increase of the dark current \citep{2022NIMPA104467510Z}. In this work, we report the characterisation of the in-orbit ageing of S13360-3050 PE multi-pixel photon counters (MPPCs) produced by Hamamatsu Photonics K.K.\footnote{\url{https://www.hamamatsu.com/jp/en.html}} using the measurements acquired by the GRBAlpha and VZLUSAT-2 CubeSat missions at LEO spanning over three years for the first time.

It has to be mentioned that the radiation-damaged SiPMs can be greatly recovered by increasing their temperature, so-called annealing \citep{2023NIMPA104867934D, 2023NIMPA105368381G, 2021NIMPA.98664673H}. The long-term annealing of SiPMs also occurs at room temperature. If a gamma-ray detector on a CubeSat at LEO is not actively cooled the SiPMs can reach this temperature. Moreover, the altitude of CubeSats can decrease significantly over a few years at LEO due to orbital decay which leads to the change of radiation conditions \cite{2020SPIE11444E..3PR}. Therefore, it is interesting to study the behaviour of SiPMs over a long time in orbit.

We add that a recent experiment has shown that it is possible to significantly recover the energy threshold of radiation-damaged SiPMs by using a current waveform amplifier, by a coincidence sum of the signal from two SiPMs attached to the same scintillator and operating at low temperature \citep{2024NIMPA106169097M}.

This article is organized as follows: Sec.~\ref{sec:cubesats} introduces the GRBAlpha and VZLUSAT-2 CubeSats and their GRB detectors. Sec.~\ref{sec:methods} describes the used methods. Sec.~\ref{sec:results} presents the obtained results, Sec.~\ref{sec:discussion} discusses the results
and Sec.~\ref{sec:conclusions} summarises the main conclusions of this work.

\section{GRBAlpha and VZLUSAT-2 CubeSats}
\label{sec:cubesats}

\begin{figure*}[h!]
	\centering
	\includegraphics[height=5.5cm]{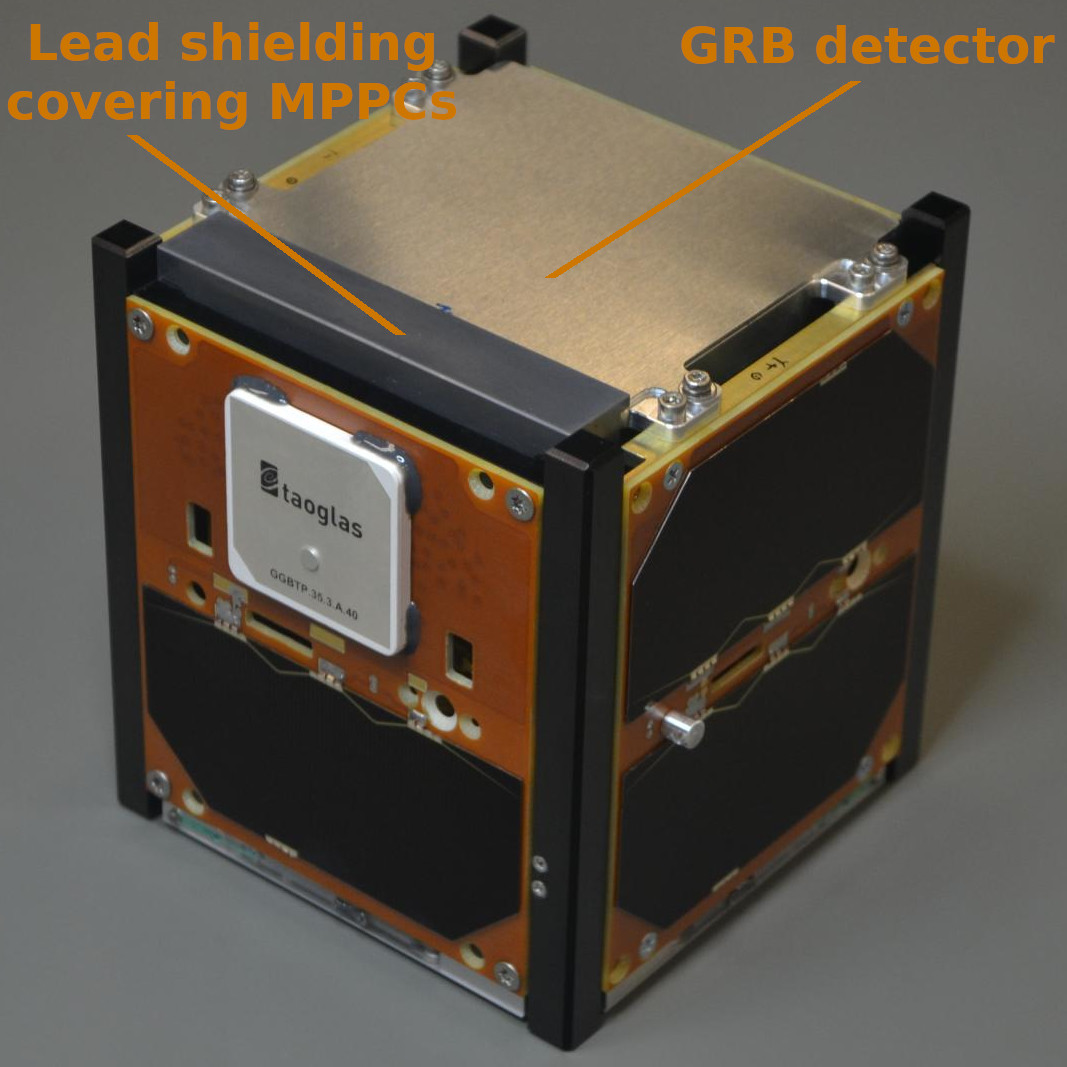}
        \includegraphics[height=5.5cm]{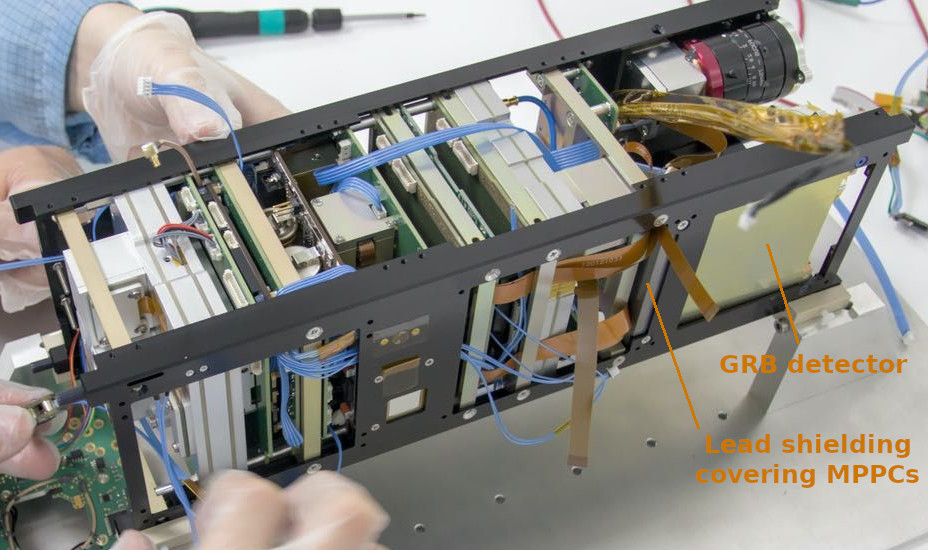}
	\caption{Left: GRBAlpha with marked GRB detector and position of the lead shielding protecting MPPCs (adopted with modification from \cite{2020SPIE11444E..4VP}). Right: VZLUSAT-2 during integration with marked one of the two GRB detectors and position of the lead shielding covering MPPCs. The second GRB detector on VZLUSAT-2 is placed at the bottom and is not visible in this photo (modified photo provided by the Czech Aerospace Research Centre).}
	\label{fig:grbalpha_vzlusat2}
\end{figure*}

GRBAlpha\footnote{\url{https://grbalpha.konkoly.hu}} is a 1U CubeSat launched on March 22, 2021, to a 550\,km altitude sun-synchronous polar orbit (SSO) \citep{2023A&A...677A..40P, 2020SPIE11444E..4VP} detecting hard X-ray / gamma-ray transients such as GRBs, soft-gamma repeaters and solar flares. The detector consists of a $75\times75\times5\,{\rm mm}$ CsI(Tl) scintillator read out by eight MPPCs S13360-3050 PE by Hamamatsu, arranged into two independent readout channels with four MPPCs connected in parallel in each readout channel. The sensitivity range covered $\sim30-900$\,keV at the beginning of the mission.

\begin{figure}[h!]
    \centering
    \includegraphics[height=5.5cm]{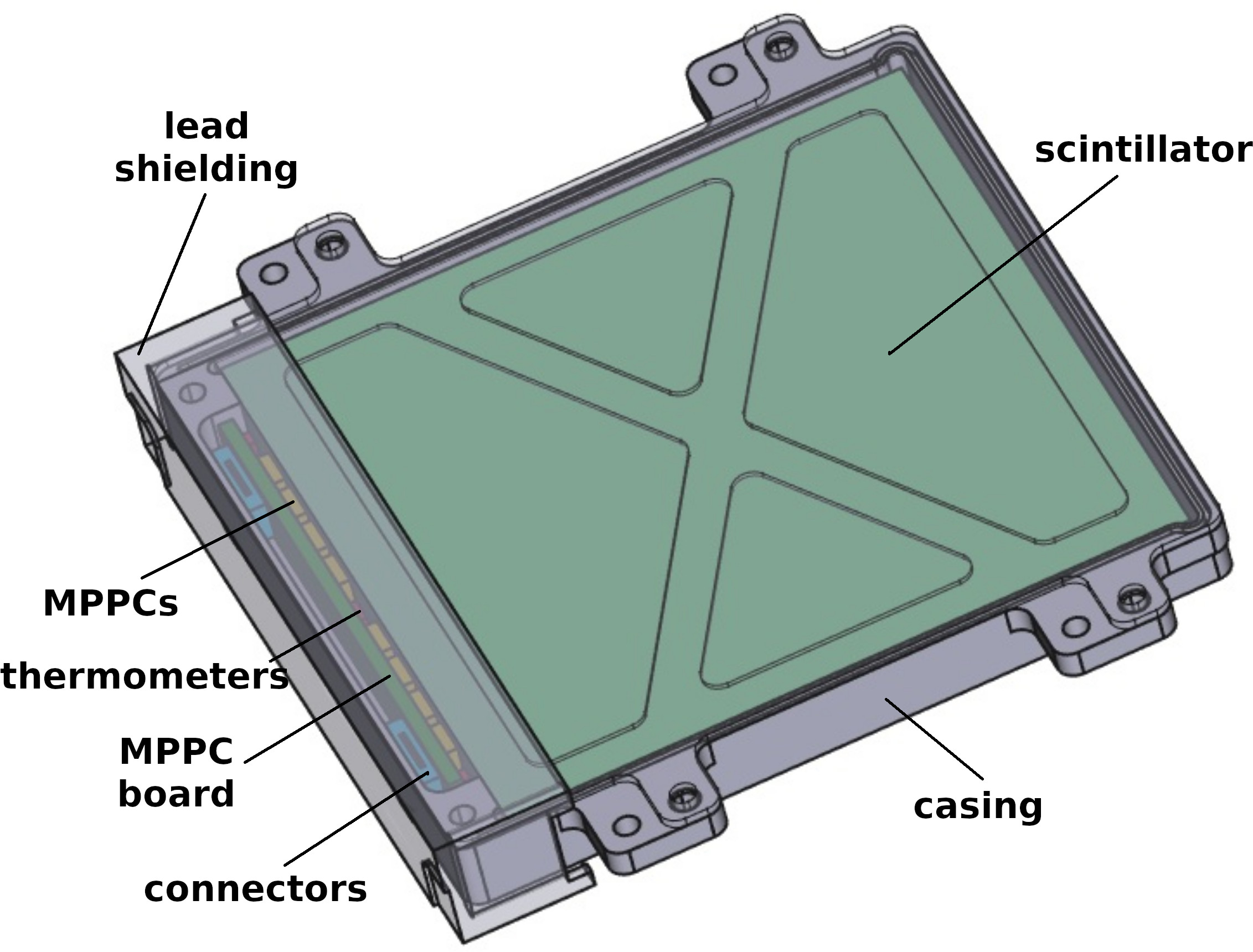}
    \caption{A diagram of the GRBAlpha's detector structure showing the main components: PbSb3 alloy lead shielding, CsI(Tl) scintillator, Al detector casing, eight MPPCs, three thermometers, MPPC board and two connectors.}
    \label{fig:grbalpha_detector}
\end{figure}

VZLUSAT-2\footnote{\url{https://www.vzlusat2.cz/en/}} is a 3U CubeSat, developed by the Czech Aerospace Research Centre (VZLU), the national centre for aerospace research, with two Earth-observing cameras as the primary payload \citep{2022Univ....8..241G, 2020SPIE11530E..0ZD}. It was launched on January 13, 2022 to a 535\,km altitude SSO. The CubeSat has multiple secondary payloads (X-ray and particle sensors) including two perpendicularly placed GRB detectors, each of them almost identical to the one on GRBAlpha.
The only two differences between the GRB detectors on both CubeSats are the following: on VZLUSAT-2 the MPPCs are coupled to the scintillator by an optical bonding pad made of the EJ-560 silicone rubber; in contrast, on GRBAlpha it is done by the DOWSIL 93-500 space grade optical glue. On VZLUSAT-2 the analog frontend electronics boards are enclosed in the detector aluminium casing. In contrast, on GRBAlpha the analog boards are placed outside the detector casing.

The gamma-ray detectors on board GRBAlpha and VZLUSAT-2 are technology demonstrations of larger detectors which could be used by future constellations of CubeSats monitoring the entire sky and localizing GRBs by using a timing-based cross-correlation technique \citep{2022SPIE12181E..1LM,2021RMxAC..53..180M,2021JATIS...7b8004G,2020SPIE11454E..1ZO,2019NIMPA.924..316T,2018SPIE10699E..2PW,2018SPIE10699E..64O,2018arXiv180603685P}.

With about 300 detections\footnote{\url{https://monoceros.physics.muni.cz/hea/GRBAlpha/}}$^,$\footnote{\url{https://monoceros.physics.muni.cz/hea/VZLUSAT-2/}} these missions have demonstrated that CubeSats can be used to monitor gamma-ray transients routinely \citep{2023A&A...677L...2R, 2022SPIE12181E..1KR}.

As mentioned in the previous section it is well known that SiPMs are prone to radiation damage, therefore, before the launch of GRBAlpha, we performed a series of simulations to investigate the possibility of partial shielding of the particle radiation at LEO to decrease the total ionizing dose (TID) in silicon in SiPMs.
Details are published in work \citep{2019AN....340..666R}, here we provide only a brief description. We considered a silicon sphere shielded by combinations of various thicknesses of Pb up to 3\,mm and 1\,mm of Al and we employed the MUlti-LAyered Shielding SImulation Software (MULASSIS)\footnote{\url{https://essr.esa.int/project/mulassis}} \citep{2002ITNS...49.2788L} integrated into the European Space Agency (ESA) SPace ENVironment Information System (SPENVIS)\footnote{\url{https://www.spenvis.oma.be}}. It employs Geant4 toolkit for the Monte Carlo simulation of the interaction of radiation with matter \citep{2016NIMPA.835..186A}. Two LEO orbits at 500\,km altitude with inclinations of $53^\circ$ and $90^\circ$ were simulated and the following proton background components were applied: geomagnetically trapped p$^+$ in Van Allen radiation belts, galactic cosmic rays (p$^+$, $^2$H, $^3$He, $^4$He), secondary proton radiation originating in Earth's atmosphere and solar protons.
We concluded that within the range of simulated shielding material the higher the thickness of Pb, the lower TID, however
strict volume and mass constraints for a CubeSat mission allowed us to use only Pb shied of the thickness of $2-2.5$\,mm.

\begin{figure*}[h!]
	\centering
        \includegraphics[angle=0,width=0.43\linewidth]{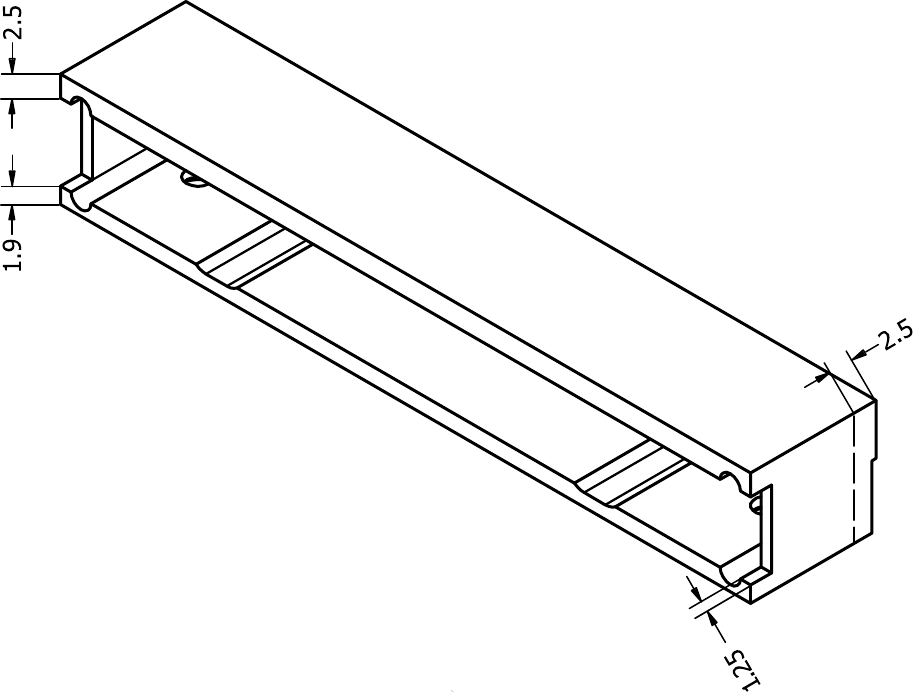}\hfill
        \includegraphics[angle=0,width=0.43\linewidth]{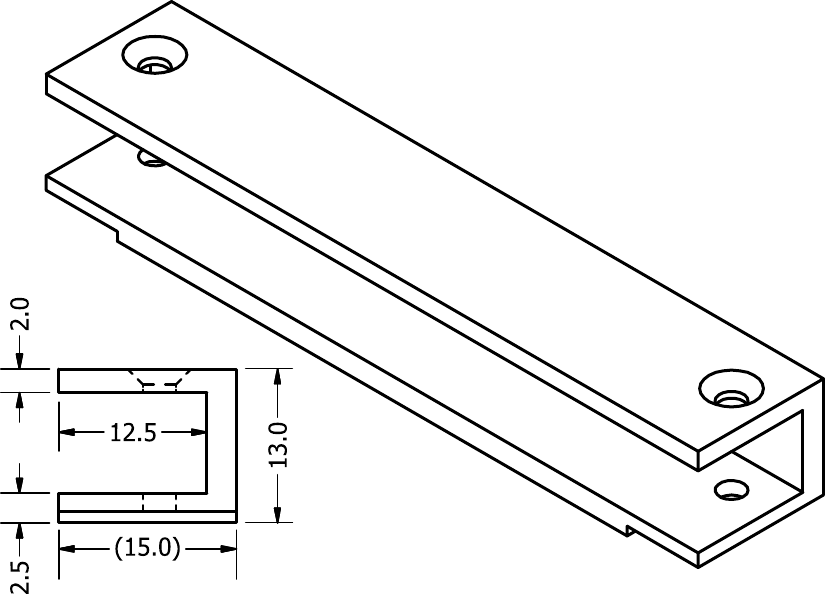}
	\caption{Left: Drawing of the lead shielding used on GRBAlpha. The shielding is made of the PbSb3 alloy. Right: Drawing of the lead shielding used on VZLUSAT-2. The shielding is made of the PbSb4 alloy. Dimensions are in millimetres.}
	\label{fig:lead_shields}
\end{figure*}

To have a better conception of GRBAlpha, VZLUSAT-2 and placement of the on-board GRB detectors, Fig.~\ref{fig:grbalpha_vzlusat2} depicts both CubeSats with detectors and the lead shielding protecting the MPPCs. Fig.~\ref{fig:grbalpha_detector} displays a diagram showing the main components of the GRBAlpha's detector. Fig.~\ref{fig:lead_shields} shows drawings of the lead shields used on both CubeSats depicting their precise shapes.

\section{Methods}
\label{sec:methods}

\subsection{Low-Energy Sensitivity Threshold and Dark Count Rate}
\label{sec:methods_thr_dcr}

In order to investigate the in-orbit degradation rate of the low-energy threshold of the gamma-ray detector system and the dark count rate of MPPCs on GRBAlpha and VZLUSAT-2 we proceed in the following way.

We regularly collect background spectra in low-background regions outside Van Allen radiation belts \citep{2020SPIE11444E..3PR} with the highest possible spectral resolution for our instruments, i.e. 256 ADC channels with the nominal operating (bias) voltage of MPPCs ($\sim 55$\,V) and typically collected over 60\,s. Due to different thermal conditions along the orbit these measurements were collected by GRBAlpha with MPPC board temperature varying from $-1^\circ$C to $+17^\circ$C. In the case of VZLUSAT-2 the MPPC board temperature varied from $-7^\circ$C to $+10^\circ$C. This background spectra collection is done by scheduling the beginning of the observation at a particular time for which the predicted position of the satellite is at the desired location in orbit. The position of the satellite for a given time was calculated from the two-line element set (TLE) obtained daily from CelesTrak\footnote{\url{https://celestrak.org}}. TLE is a data format containing a list of orbital elements for a given time commonly used to calculate the position of a satellite.

\begin{figure*}[h!]
	\centering
 	\includegraphics[trim={-2.0cm 0 0 0},clip=true,angle=270,width=0.49\linewidth]{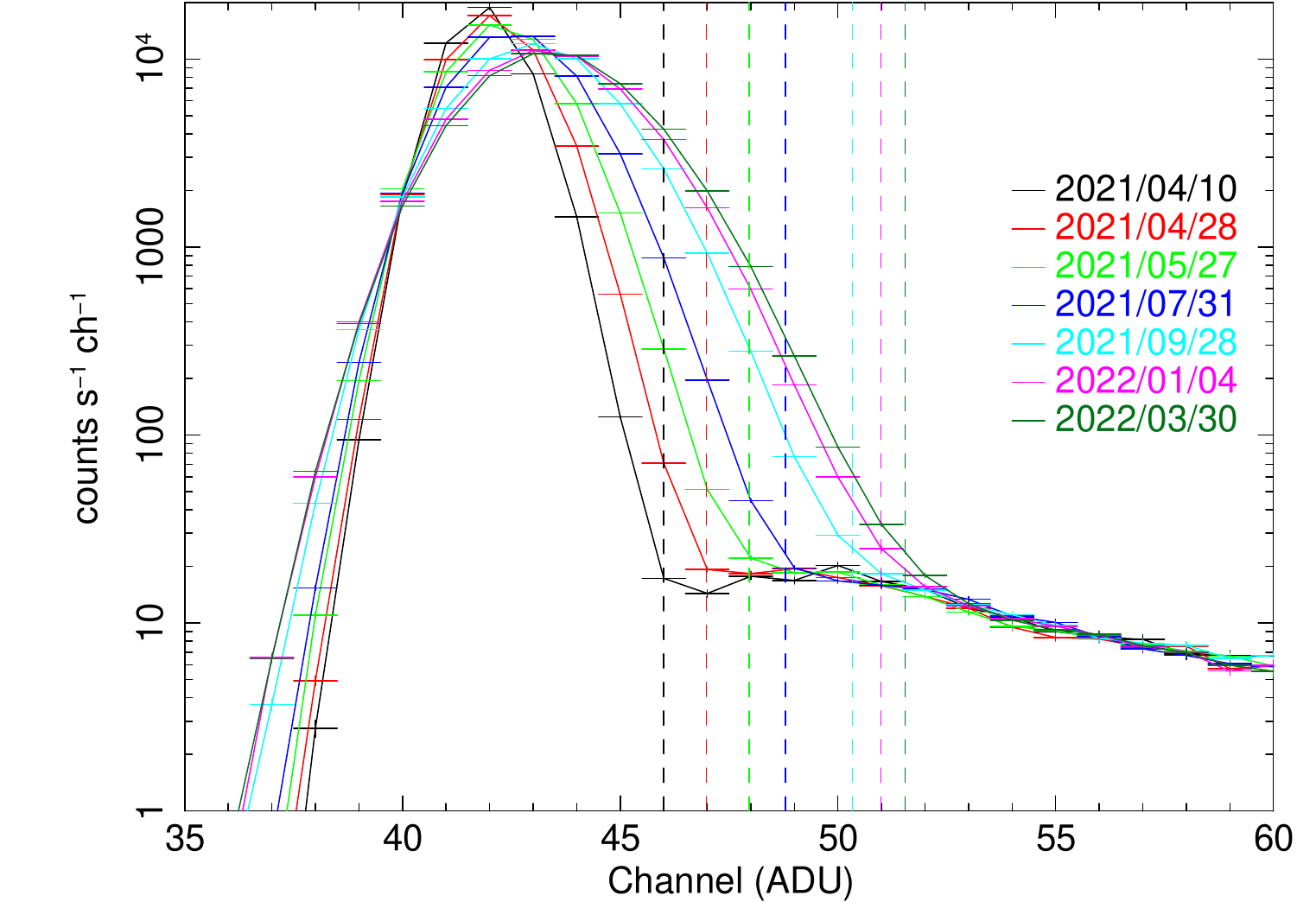}
    \includegraphics[angle=270,width=0.49\linewidth]{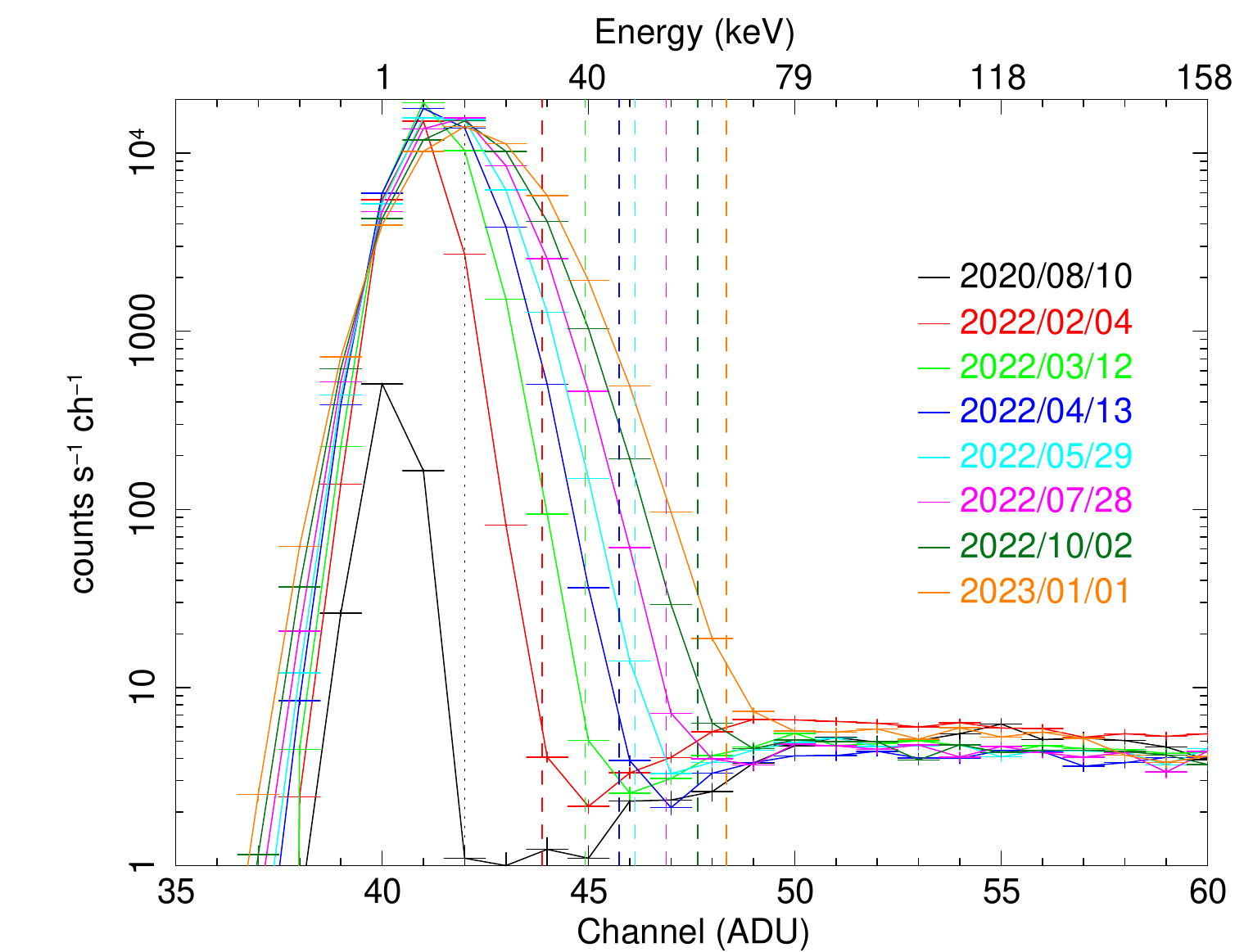}
	\caption{Left: Measurement of the noise peak evolution observed in the background spectrum of readout channel \emph{ch0} of the GRBAlpha detector over the first year on orbit. Right: The noise peak spectrum evolution measured by the VZLUSAT-2 GRB detector unit 1 (readout channel \emph{ch2}) over the first year on orbit. The spectrum in black colour (2020/08/10) was measured during the ground calibrations. The vertical dashed lines in both panels mark the low energy thresholds. The black dotted vertical line marks a pre-launch threshold for VZLUSAT-2 and it is shown only for reference because it was not calculated with the method employing a template spectrum based on the post-launch measurement.}
	\label{fig:noise_peak_spec_grbalpha}
\end{figure*}

The low-energy part of a measured background spectrum is dominated by the dark noise peak caused by thermal fluctuations of charge carriers in MPPC pixels which trigger avalanches. Fig.~\ref{fig:noise_peak_spec_grbalpha} presents the noise peak evolution observed in the background spectra of the GRBAlpha and VZLUSAT-2 GRB detectors over the first year on orbit.

\begin{figure*}[h!]
	\centering
	\includegraphics[angle=0,width=0.49\linewidth]{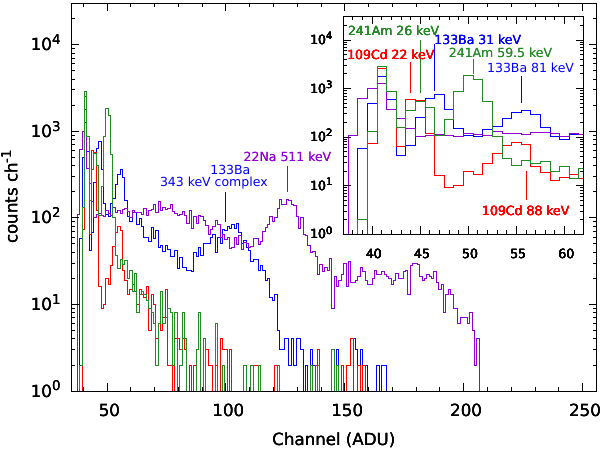}
    \includegraphics[angle=0,width=0.49\linewidth]{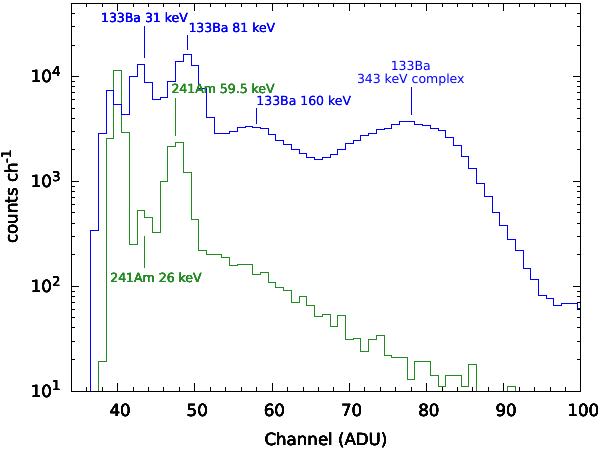}
	\caption{Laboratory measurements of gamma-ray lines of radioisotopes used for the gain calibration. Left: The calibration spectra recorded by GRBAlpha's detector readout channel \emph{ch0} on 2020/12/04. Right: The calibration spectra recorded by VZLUSAT-2's detector unit 0 readout channel \emph{ch0} on 2020/08/10.}
	\label{fig:calibration_spectra}
\end{figure*}

To determine the upper boundary of the dark noise peak which defines the low-energy threshold of the sensitivity of our detector, we employ a method based on subtracting a template spectrum $S_\mathrm{t}$ from the actual measured spectrum $S$. This method is described in the following steps:

\begin{enumerate}
    \item Define a template spectrum for each readout channel of detectors on GRBAlpha and VZLUSAT-2 separately. Each readout channel is composed of four MPPCs connected in parallel. In what follows the readout channels for GRBAlpha are marked as \emph{ch0} and \emph{ch1}. For VZLUSAT-2 the readout channels are marked as \emph{ch0} and \emph{ch2} on detector units 0 and 1, respectively.
    A template spectrum $S_\mathrm{t}$ was constructed from a low-background spectrum measured in orbit as soon as possible after the launch of the CubeSat when the exposed dose in MPPCs was still minimal. We present the simulation results in Sec~\ref{sec:res-thr}. For GRBAlpha and both readout channels, these spectra were measured on 2021/04/12, i.e. 21 days after the launch and integrated over 150\,s. For VZLUSAT-2 and both readout channels, these spectra were measured on 2022/02/04, i.e. 22 days after the launch and integrated over 60\,s. In the template spectrum $S_\mathrm{t}$ the dark noise peak is replaced by a constant value equal to the count rate of the first spectral channel measured immediately on the right side from the noise peak, see the left panel of Fig.~\ref{fig:method}.
    
    \item Create the difference spectrum $S_\mathrm{d} = S - S_\mathrm{t}$ between the actual measured spectrum $S$, for which we want to determine the noise peak threshold channel $x_\mathrm{thr}$, and the template spectrum $S_\mathrm{t}$.
    
    \item While looking at the difference spectrum $S_\mathrm{d}$ from the lowest spectral channels to the highest channels on the right side from the noise peak, we find the spectral channel $x_2$ where the difference spectrum falls below a predefined rate level $y_\mathrm{thr}=10$\,cnt/s. In other words, we look for a spectral channel $x_2$ where the measured spectrum $S$ starts to be closer to the template spectrum $S_\mathrm{t}$ by 10\,cnt/s or less. Then we define a channel $x_1$ as the nearest channel lower than $x_2$, i.e. $x_1=x_2-1$. Next, we denote $y_1=S_\mathrm{d}(x_1)$ and $y_2=S_\mathrm{d}(x_2)$ as marked in the right panel of Fig.~\ref{fig:method}.

\begin{figure*}[h!]
	\centering
	\includegraphics[angle=270,width=0.49\linewidth,trim={-1.0cm 0 0 0},clip]{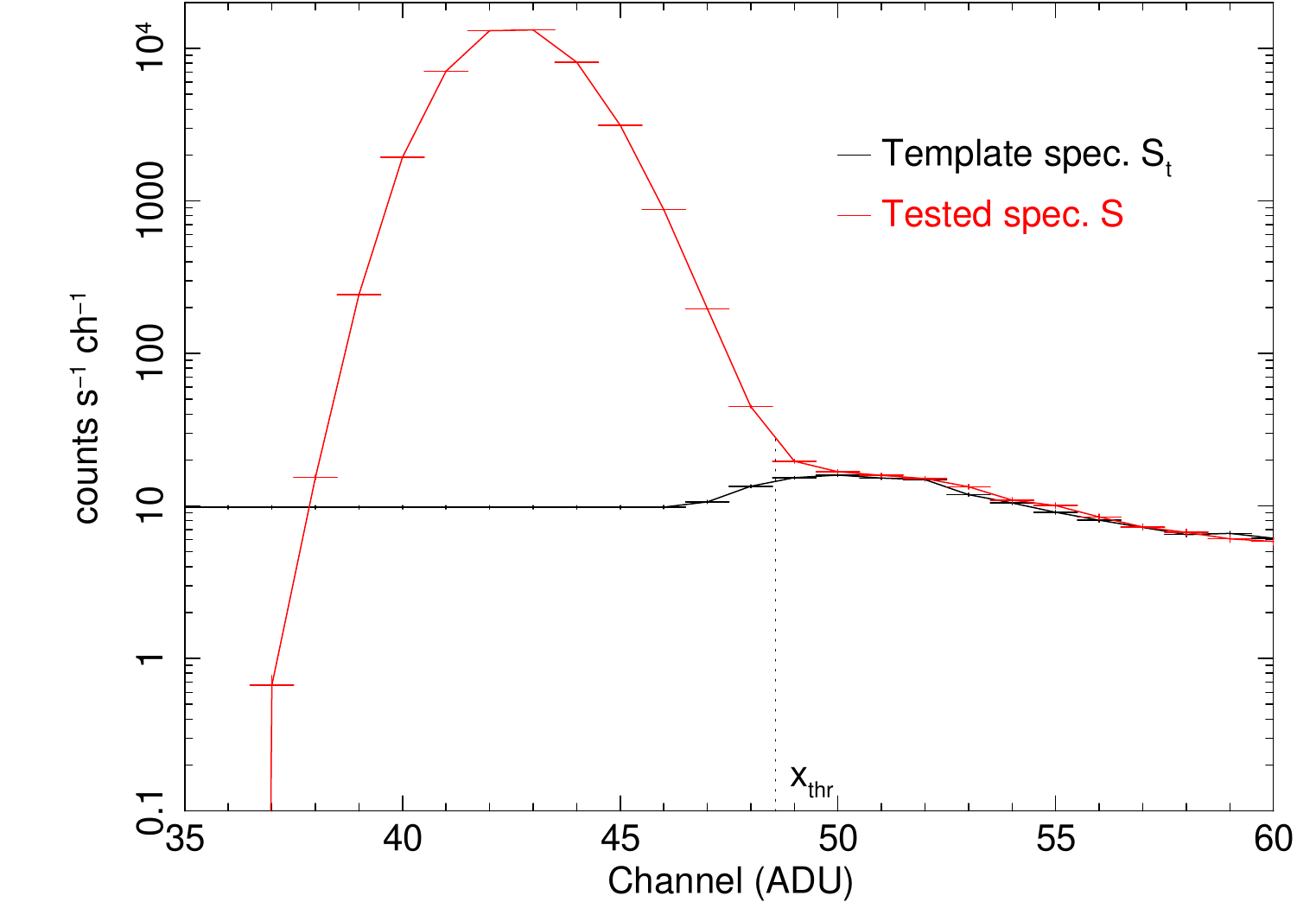}
    \includegraphics[angle=270,width=0.49\linewidth]{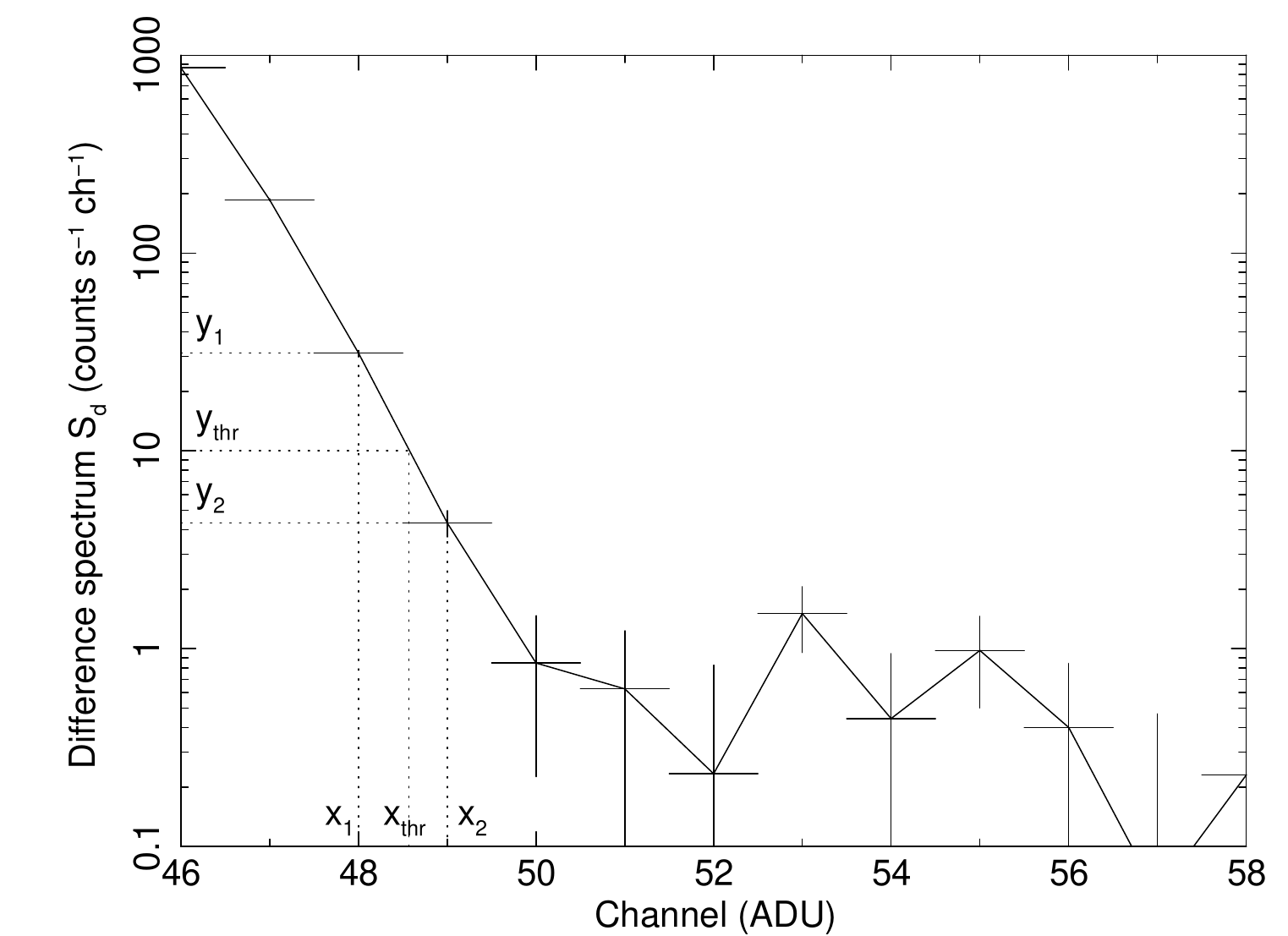}
	\caption{Left: The black curve shows the template spectrum $S_\mathrm{t}$ for the GRBAlpha readout channel \emph{ch0}. The red curve shows an example of a background spectrum $S$ measured on orbit by GRBAlpha, the readout channel \emph{ch0}, on 2021/07/31. Right: The difference spectrum $S_\mathrm{d}$ with marked channels $x_1$ and $x_2$ right below and above the calculated threshold channel $x_\mathrm{thr}$ with corresponding $y$ values.}
	\label{fig:method}
\end{figure*}
    
    \item Given the values of $x_1$, $x_2$, $y_1$, $y_2$ and $y_\mathrm{thr}$ one can calculate the value of the noise peak threshold channel $x_\mathrm{thr}$ with sub-ADU resolution from a linear interpolation:
    \begin{equation}\label{eqn:xthr}
    x_\mathrm{thr}=(x_1-x_2)(y_\mathrm{thr}-y_2)/(y_1-y_2) + x_2\hspace{0.5em}
    \end{equation}
    
    \item In order to obtain the noise peak energy threshold $E_\mathrm{thr}$ from the threshold channel $x_\mathrm{thr}$ one has to apply gain calibration. For the pre-launch calibration of the GRBAlpha's detector we use a linear gain calibration based on laboratory measurements performed at room temperature by measuring gamma-ray lines of radioisotopes $^{241}$Am (26\,keV, 59.5\,keV), $^{133}$Ba (31\,keV, 81\,keV), $^{109}$Cd (22\,keV, 88\,keV) and $^{22}$Na (511\,keV).
    The lines were fitted by a Gaussian function, Gaussian plus constant, Gaussian plus linear function or Gaussian plus a parabola and the fit with the lowest reduced $\chi^2_\nu$ was used to determine the position of the line. The best-fit value of the mean of the Gaussian and its $1\sigma$ uncertainty for each line was taken as a result.
    For the example of the calibration spectra see the left panel of Fig.~\ref{fig:calibration_spectra} with the measurements from the readout channel \emph{ch0}. The calibration spectra for both readout channels were recorded with MPPCs' operating voltage setting of 168\,ADU. During the nominal operations of the satellite in orbit, we use a slightly higher operating voltage setting of MPPCs of 178\,ADU in order to increase the gain of the detector. The results of the laboratory gain calibration measurements for different operating voltage settings were extrapolated to 178\,ADU and one linear gain calibration function is used for both readout channels as follows:
    \begin{equation}\label{eqn:calib}
    E=g\,x+E_0,
    \end{equation}
    where $E$(keV) is the inferred energy, $x$(ADU) is the spectral channel number (counted from 1), $g=4.08$\,keV/ADU is the gain factor and $E_0=-154$\,keV is the offset.
    
    We noticed that after operating the satellite in orbit for more than three years, there is a substantial change in the detector's gain. Fortunately, we have enough regularly collected measurements for the activation lines study, which allows us to analyse the long-term gain evolution and recalibrate the detector. Special data collection for the activation lines study was composed of scheduling the measurements to record sequences of high-resolution spectra (256 ADC channels) with 60\,s exposures during and after a passage through the South Atlantic Anomaly (SAA) when the satellite moves from south to north towards the low-background region. The measurements were done with nominal operating (bias) voltage of MPPCs (setting of 178 DAU) as well as with lower operating voltages (setting of 160, 155 and 150 DAU). The spectra revealed two activation lines. The spectra collected with the nominal operating voltage revealed an activation line associated with the energy of 260\,keV. For lower operating voltage of MPPCs we observe, besides the 260\,keV also the 511\,keV electron-positron annihilation line. Both lines vanish within a few minutes after the SAA passage. We stacked the individual spectra which exhibit the activation lines to increase the signal-to-noise ratio. Fig.~\ref{fig:activ_line_spectra} shows the spectra with the observed activation lines. Next, we fitted the stacked spectra with a Gaussian function, Gaussian plus constant, Gaussian plus linear function or Gaussian plus parabola and determined the position of a line as the mean of the Gaussian with its $1\sigma$ uncertainty. The left panel of Fig.~\ref{fig:grbalpha_gain_evolution} displays the dependence of the position of the 260\,keV activation line on the number of days the satellite spent in orbit. Taking the Eq.~(\ref{eqn:calib}) and assuming the constant offset $E_0$ equal to the pre-launch value one can calculate the gain factor $g$ as a function of the number of days since launch. The right panel of Fig.~\ref{fig:grbalpha_gain_evolution} displays that dependence. We find the best with by a cubic function as:
    \begin{equation}\label{eqn:gain_time}
    g(t) = 5.54\times10^{-10}t^3 – 1.91\times10^{-6}t^2 + 2.29\times10^{-3}t + 4.02,
    \end{equation}
    where $t$ is the number of days since launch. Therefore, for the in-orbit data of GRBAlpha we use the calibration formula:
    \begin{equation}\label{eqn:in-orbit_calib}
    E=g(t)\,x+E_0,
    \end{equation}
    where $E_0$ has the fixed pre-launch value of $-154$\,keV and the gain factor $g(t)$ is given by Eq.~(\ref{eqn:gain_time}).

    A dedicated Geant4 simulation was performed to understand the isotopes behind the line at 260\,keV. A detailed model of the GRBAlpha satellite was irradiated isotropically by protons. Two specific simulations were carried out, one with 10 million protons with 150\,MeV. The other one had the same number of primaries but with an energy of 700\,MeV. We have registered all radioactive decays within the first 1000\,s of irradiation. Then we filtered out nuclei that had an excitation energy between 230\,keV and 300\,keV and half-life between 1\,s and 100\,000\,s. The reasoning behind this is that we expect this line to originate from $\gamma$-rays that originate from the deexcitation of nuclei since at these energy levels this is the most likely source. In the case of 150\,MeV the three recorded isotopes were $^{125}$Xe[252.61] and $^{127}$Xe[297.1] in the scintillators and $^{70}$Cu[242.6] in the lead shielding. The latter had a half-life of only 6\,s so we excluded this isotope. $^{125}$Xe[252.61] and $^{127}$Xe[297.1] had similar original activity with half-life of 57\,s and 69\,s respectively. In the other simulation where the primary protons had an energy of 700\,MeV the same isotopes of Xe were by far the most abundant. Therefore, we conclude that these two isotopes are the likely source of the line around 260\,keV since the detector energy resolution is not sufficient to resolve them.
    
    In the case of VZLUSAT-2 we have only one observation of activation lines at 260\,keV and 511\,keV at the beginning of 2024. Unfortunately, we do not have enough in-orbit measurements of activation lines, which would allow us to reliably trace the in-orbit gain degradation. Therefore, in what follows we have to rely only on the pre-launch calibrations. We use a linear gain calibration from pre-launch laboratory measurements of lines from radioisotopes $^{241}$Am (59.5\,keV), $^{133}$Ba (31\,keV, 81\,keV, 160\,keV, complex of lines unresolved from each other at 276\,keV, 303\,keV, 356\,keV and 383\,keV with weighted average of 343\,keV) at room temperature, see right panel of Fig.~\ref{fig:calibration_spectra}.   
    The gain calibration for the readout \emph{ch0} (detector unit 0) is $E\mathrm{(keV)}=9.12\,x-366$\,keV. and for the readout \emph{ch2} (detector unit 1) it is $E\mathrm{(keV)}=7.84\,x-313$\,keV.
    The calibration parameters for detectors on GRBAlpha and two readout channels on VZLUSAT-2 are different due to MPPCs' mildly different operating voltages. The gain calibration curves for GRB detectors on both CubeSats are presented in Fig.~\ref{fig:calibration_curves}.
    It has to be mentioned that the pre-launch laboratory calibration measurements were performed at the constant room temperature however, work \citep{10.1117/12.3025855} shows that the calibration gain factor is a function of temperature. The gain calibration factor change can be $\sim20-30$\,\% for the temperature range of 0$^\circ$C$-20^\circ$C depending on the operating voltage of MPPCs. This was measured in the laboratory on a flight spare model of the GRB detector of the GRBBeta CubeSat. That detector is almost the same as the one on board of GRBAlpha. Therefore, ideally one should use a calibration between ADU and energy which also accounts for the detector's temperature, however for GRBAlpha and VZLUSAT-2 we do not have such calibration measurements available.

    \item The noise peak of background spectra shown in Fig.~\ref{fig:noise_peak_spec_grbalpha} and \ref{fig:method} is dominated by dark count rate. Therefore, by integrating this noise peak, one can obtain an approximation of the dark count rate (DCR). There is a very small amount of low-energy gamma-rays which pass through 1\,mm thick Al detector casing and deposit their energy in the scintillator, however, in low-background regions on orbit the count rate from these events is negligible compared to the thermal dark count rate. Therefore, we approximate the dark count rate as:
    \begin{equation}\label{eqn:dcr}
    DCR\approx\sum_{0}^{x_\mathrm{thr}}S(ADU)
    \end{equation}

\end{enumerate}

\begin{figure*}[h!]
	\centering
	\includegraphics[angle=0,width=0.49\linewidth]{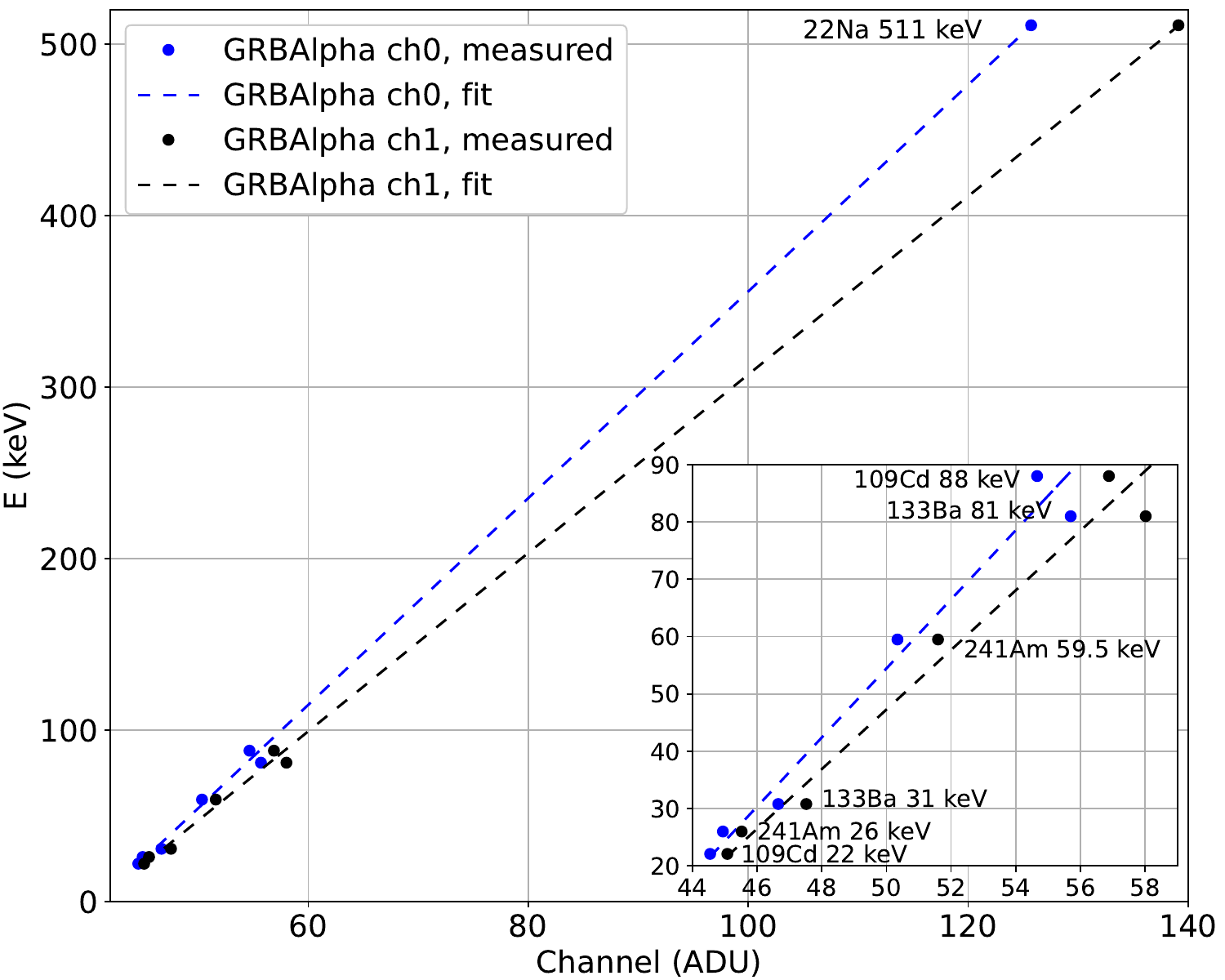}
    \includegraphics[angle=0,width=0.49\linewidth]{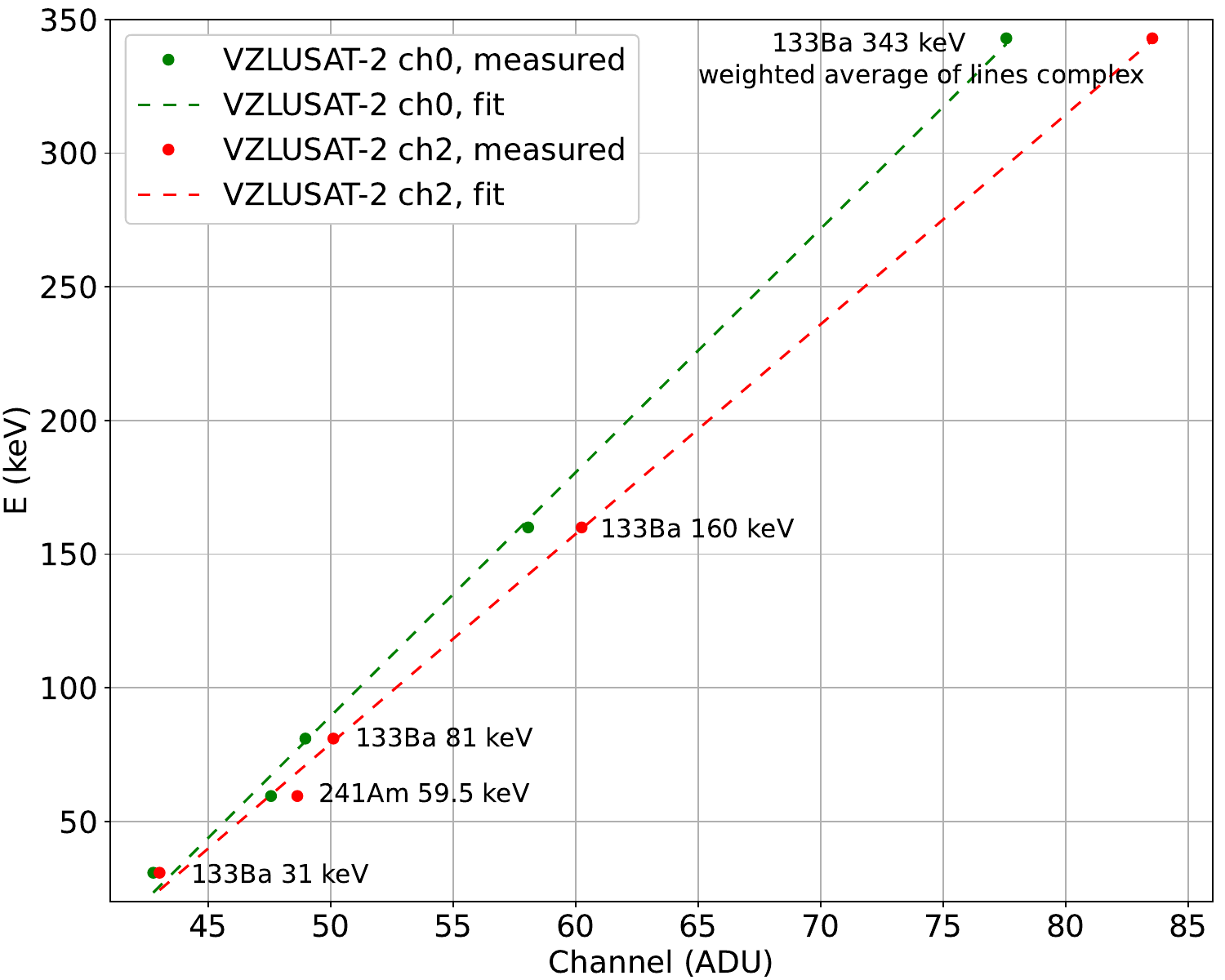}
	\caption{The gain calibration curves obtained from the laboratory measurements of the GRB detectors on GRBAlpha (left) and on VZLUSAT-2 (right). The $1\sigma$ uncertainties of the measured position of each line are lower than 0.2\,ADU and are too small to be visible in the figure.}
	\label{fig:calibration_curves}
\end{figure*}

\begin{figure*}[h!]
	\centering
	\includegraphics[width=0.49\linewidth]{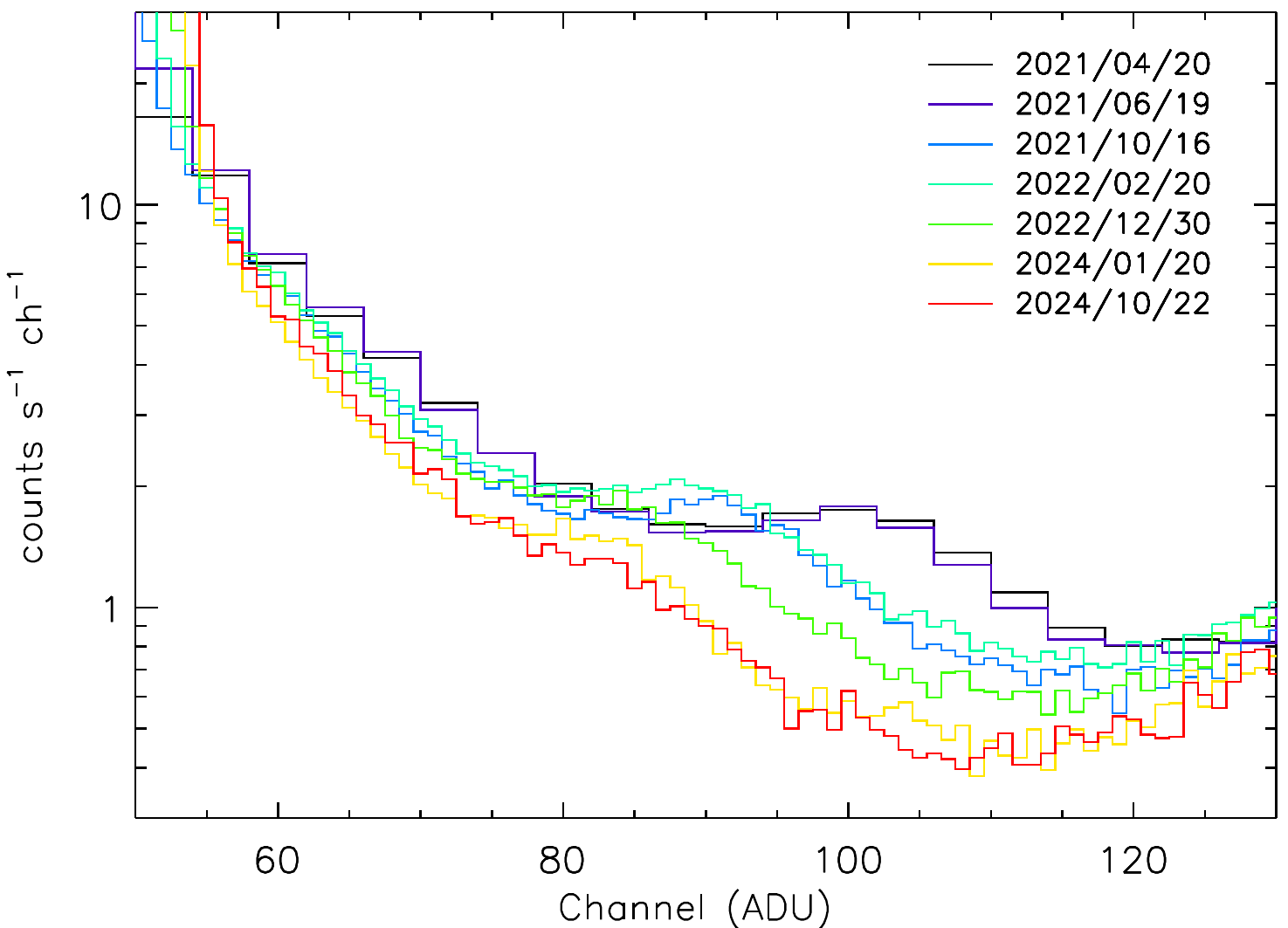}
    \includegraphics[width=0.49\linewidth]{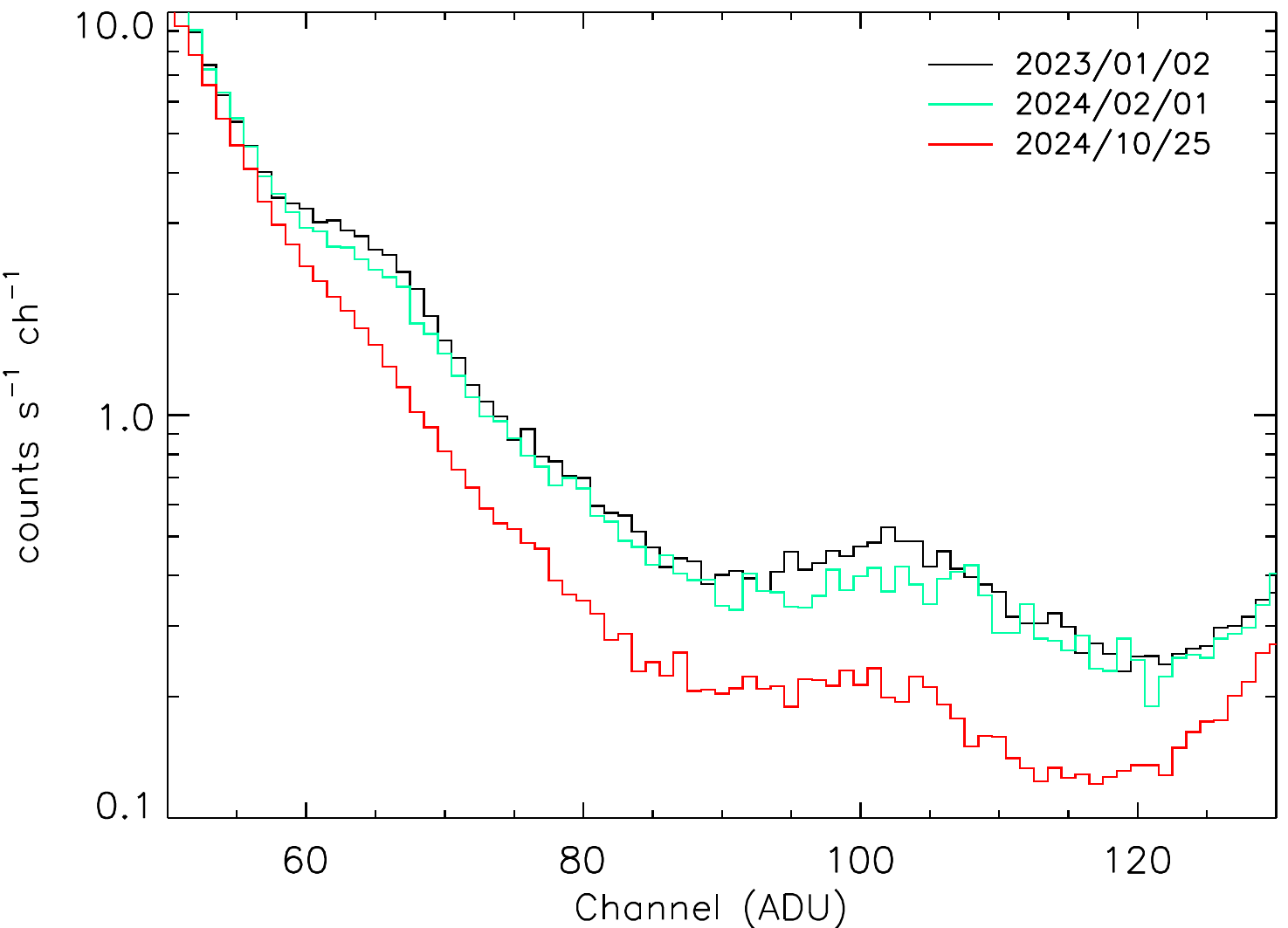}
	\caption{Stacked spectra with activation lines observed by GRBAlpha with the readout channel \emph{ch0} after SAA passages. Left: Displayed is the activation line associated with the energy of 260\,keV used to recalibrate the detector in orbit. The measurement was done with the nominal operating (bias) voltage of MPPCs (setting of 178\,DAU). Right: An activation line associated with the energy of 260\,keV (channel number $\sim65$\,ADU) and the 511\,keV electron-positron annihilation line (channel number $\sim100$\,ADU). The measurement was done with a lower operating voltage of MPPCs (setting of 150\,DAU). In both panels, the increase of the spectra at the highest channels is likely due to the non-linearity of the front-end analogue electronics.}
    \label{fig:activ_line_spectra}
\end{figure*}

\begin{figure*}[h!]
	\centering
        \includegraphics[width=0.49\linewidth]{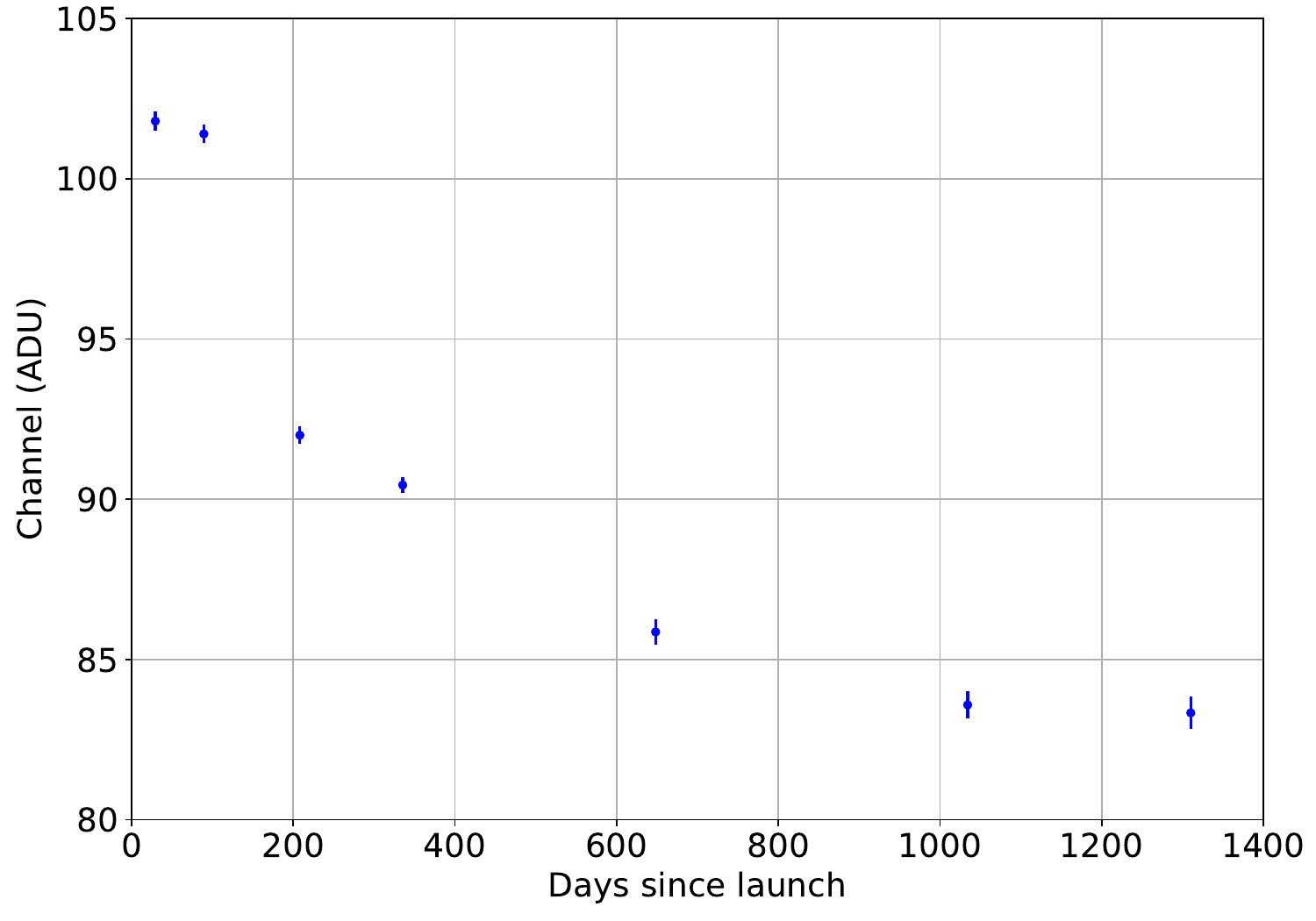}
	\includegraphics[width=0.49\linewidth]
    {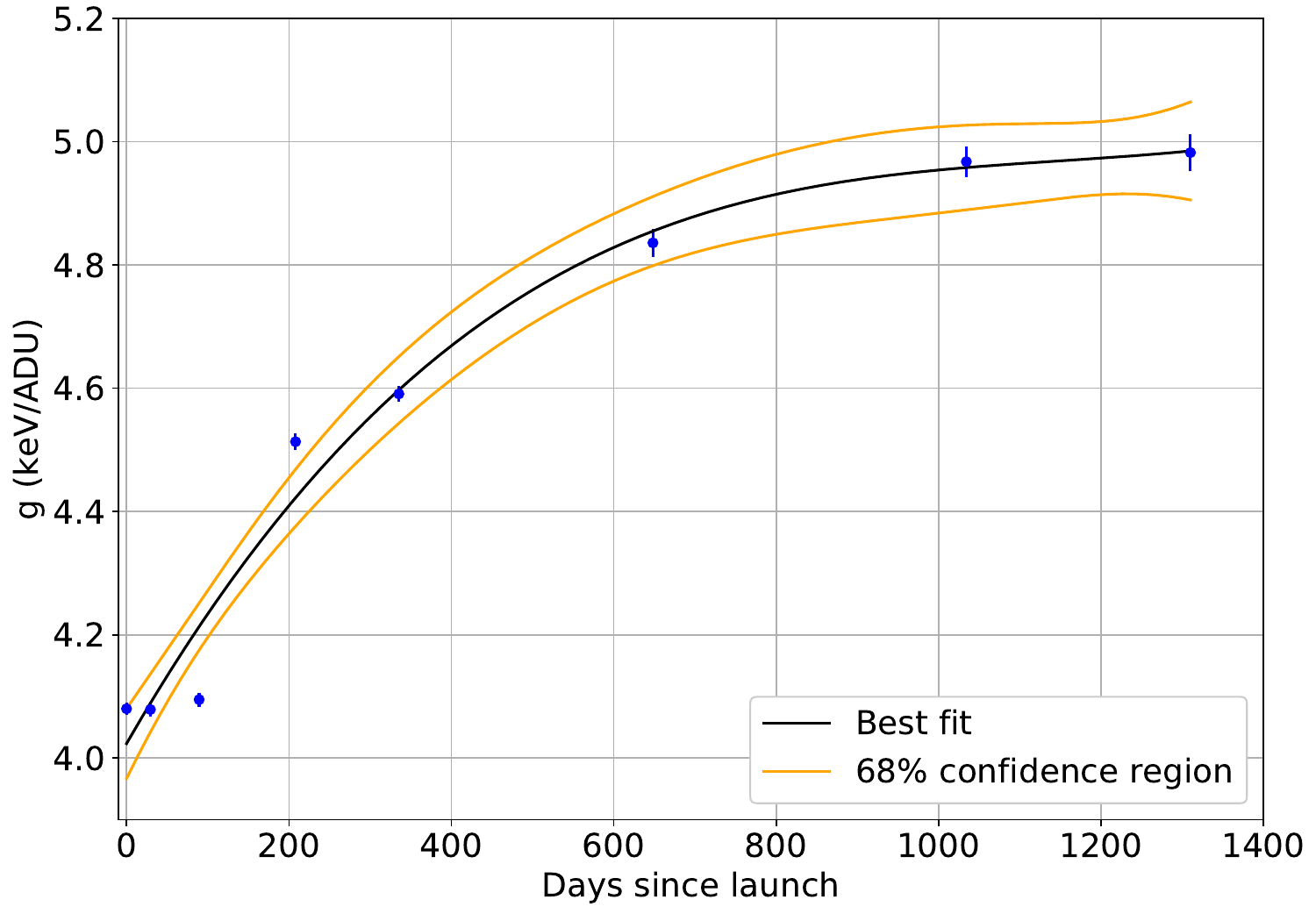}
	\caption{Left: The position of the activation line in terms of the channel number associated with 260\,keV observed by GRBAlpha with the readout channel \emph{ch0} as a function of time the satellite spent in orbit indicating the detector's gain degradation. Right: The calculated gain factor of the GRBAlpha detector as a function of time the satellite spent in orbit. The pre-launch gain factor was used for the time zero. The solid curve shows the best fit by a cubic function.}
	\label{fig:grbalpha_gain_evolution}
\end{figure*}

\subsection{Expected In-Orbit TID and TNID}
\label{sec:methods_expected_tid_nid}

To simulate the expected TID and
the total non-ionizing dose or displacement damage (TNID) \footnote{\url{https://ecss.nl/hbstms/ecss-e-hb-10-12a-calculation-of-radiation-and-its-effects-and-margin-policy-handbook/}} in
MPPCs since the launch of these CubeSats we proceed as follows. First, we used the actual TLEs of GRBAlpha and VZLUSAT-2. TLEs were obtained from the CelesTrak's satellite catalog (SATCAT). We separated the whole period of the satellite in orbit into several mission segments. Each segment contains about one-month time interval and we used individual TLE for each mission segment. Thus, we can cover the progress of the satellite orbital decay.

Second, we applied the IRENE v1.57.004 software provided by the U.S. Air Force Research Laboratory\footnote{\url{https://www.vdl.afrl.af.mil/programs/ae9ap9/}} to calculate the omnidirectional differential particle fluence of geomagnetically trapped electrons (AE-8 model) and protons (AP-8 model) \citep{2020SPIE11444E..3PR, 2005igtr.book.....W, 2014JGRA..119.9658Q} for each mission segment. We choose the condition of the solar cycle activity minimum until Aug 2022. From Sep 2022 we choose the solar cycle activity maximum.
We note that a quoted error estimate of proton flux according to the AP-8 model is of a factor of \textbf{two}
\footnote{\url{https://www.spenvis.oma.be/help/background/traprad/traprad.html}}.

\begin{figure*}[h!]
	\centering
	\includegraphics[height=4.5cm]{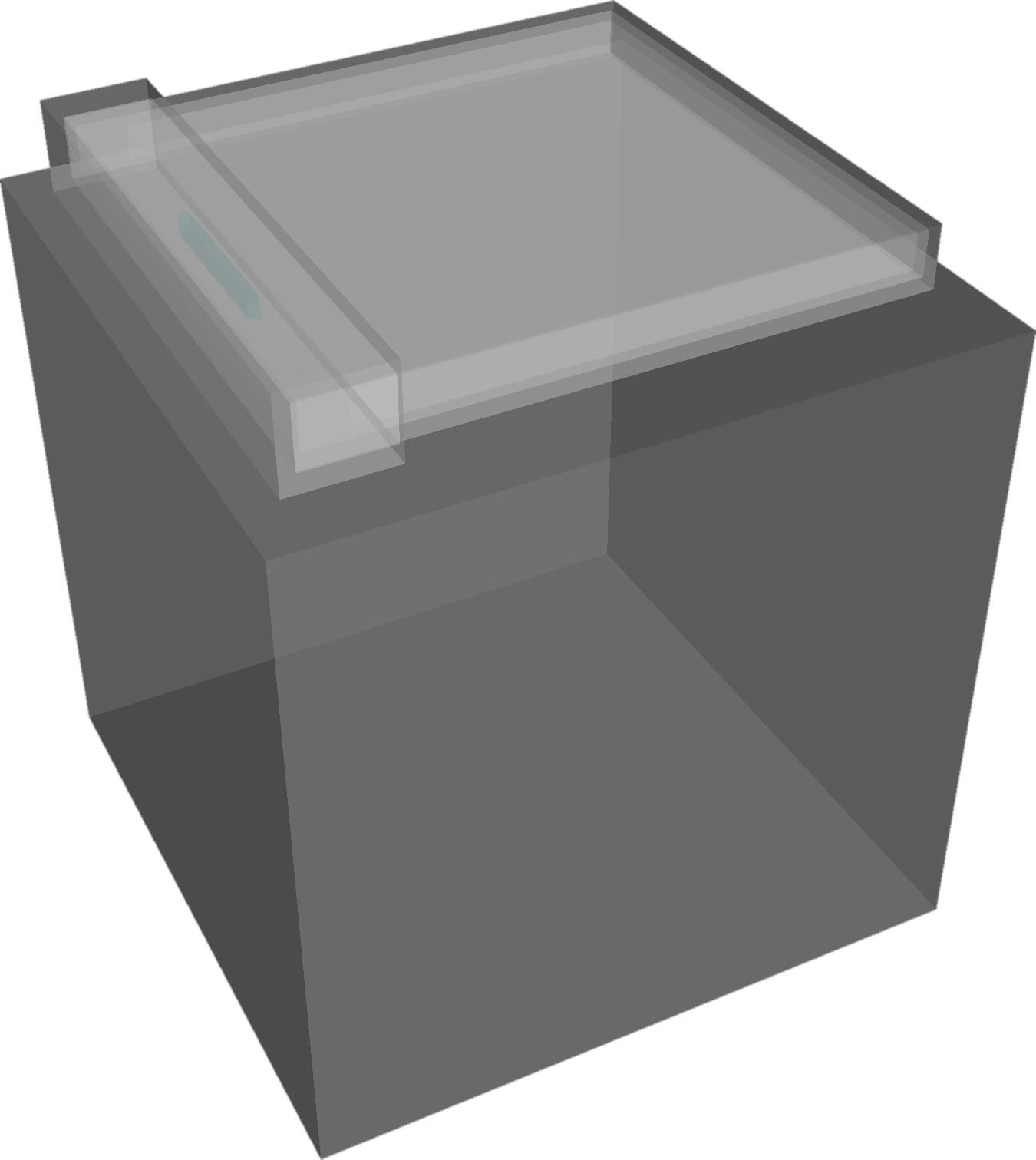}
        \hspace{1cm}
        \includegraphics[height=4.5cm]{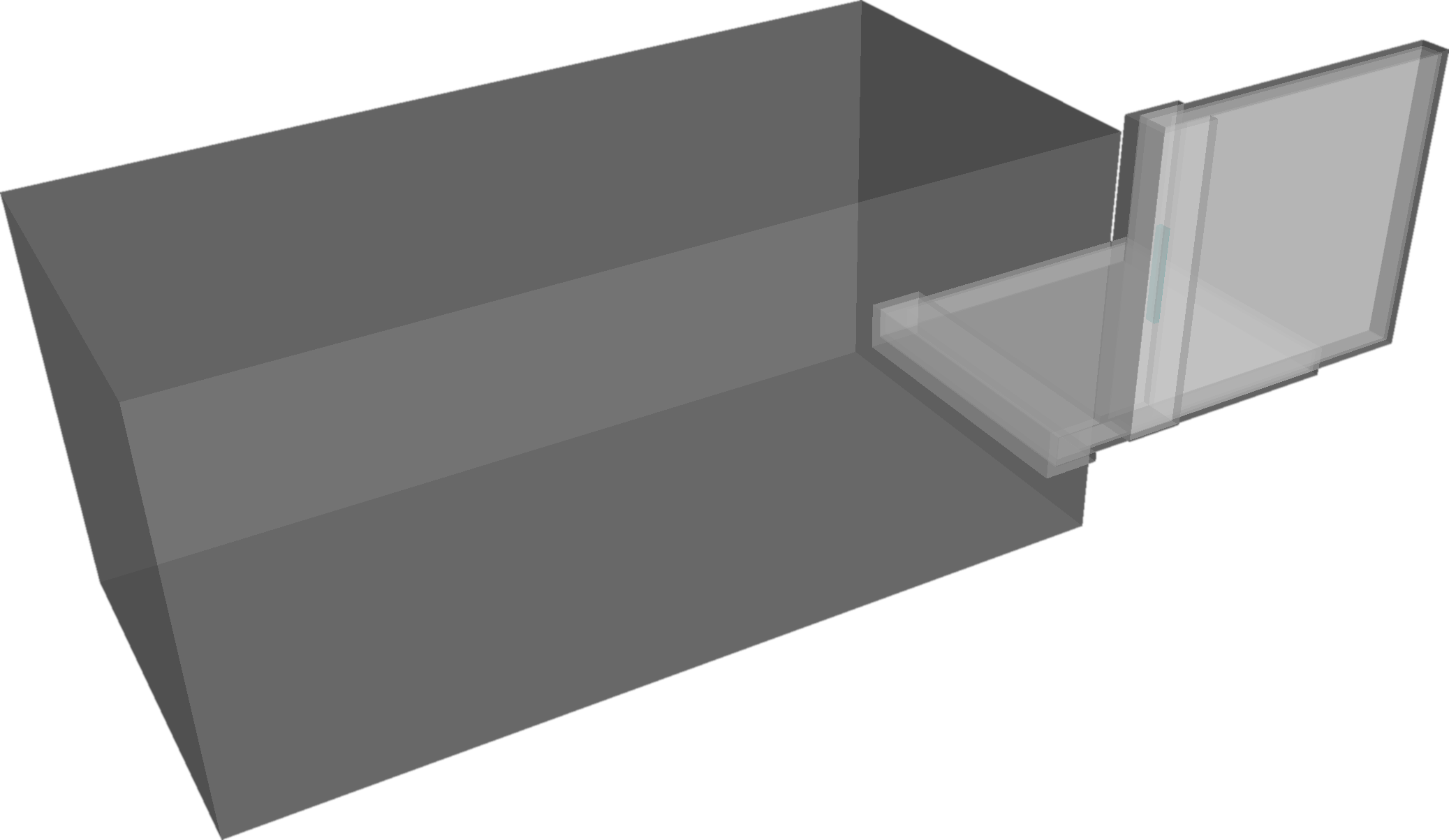}
	\caption{The mass models of GRBAlpha (left) and VZLUSAT-2 (right) CubeSats used for the calculation of TID and TNID are displayed. The models contain detailed shapes and compositions of the detectors (CsI scintillator, Al casing, Pb shield, SiPM). The satellite bus with all other sub-systems is approximated by a box of the dimensions and average density representing the mass of the rest of the satellite.}
	\label{fig:grbalpha_vzlusat2_mass_model}
\end{figure*}

Third, having the particle fluence for each mission segment we simulated the TID and TNID deposited in Si using
the ESA's Geant4 Radiation Analysis for Space (GRAS)\footnote{\url{https://space-env.esa.int/software-tools/gras}} software package \citep{2005ITNS...52.2294S}. It is a Geant4-based tool for analyses of radiation effects in materials which works with complex 3D geometry models. We used GRAS-06-00-01 with Geant4-10-07-p01. For non-ionizing energy loss coefficients in Si, we chose JPL/NRL/NASA (2003) \citep{2003ITNS...50.1919M}. Fig.~\ref{fig:grbalpha_vzlusat2_mass_model} shows the mass models of GRBAlpha and VZLUSAT-2 CubeSats used for these simulations. The models contain the detectors' detailed shapes and compositions including the CsI scintillator, Al casing, Pb shield and a block of $24\times3\times1.5$\,mm$^3$ of Si representing 8 MPPCs each of size of $3\times3\times1.5$\,mm$^3$. The satellite bus with all other sub-systems is approximated by a box of the correct dimensions and average density representing the mass of the rest of the satellite. For GRBAlpha the satellite bus is approximated by a cube of dimensions of $10\times10\times10$\,cm$^3$ and density of $0.92$\,kg/m$^3$. For VZLUSAT-2 the satellite bus is approximated by a block of dimensions of $21\times10\times10$\,cm$^3$ and density of $1.58$\,kg/m$^3$. Particles were shot towards a satellite isotropically from a surrounding sphere, centred in the middle of the detector of a radius of 15\,cm for GRBAlpha and 30\,cm for VZLUSAT-2. We simulated TID in rad units and TNID in 1\,MeV neutron equivalent fluence (cm$^{-2}$).

\section{Results}
\label{sec:results}

\subsection{Low-Energy Threshold and DCR}
\label{sec:res-thr}

\begin{figure}[h!]
	\centering
	\includegraphics[height=0.72\textheight]{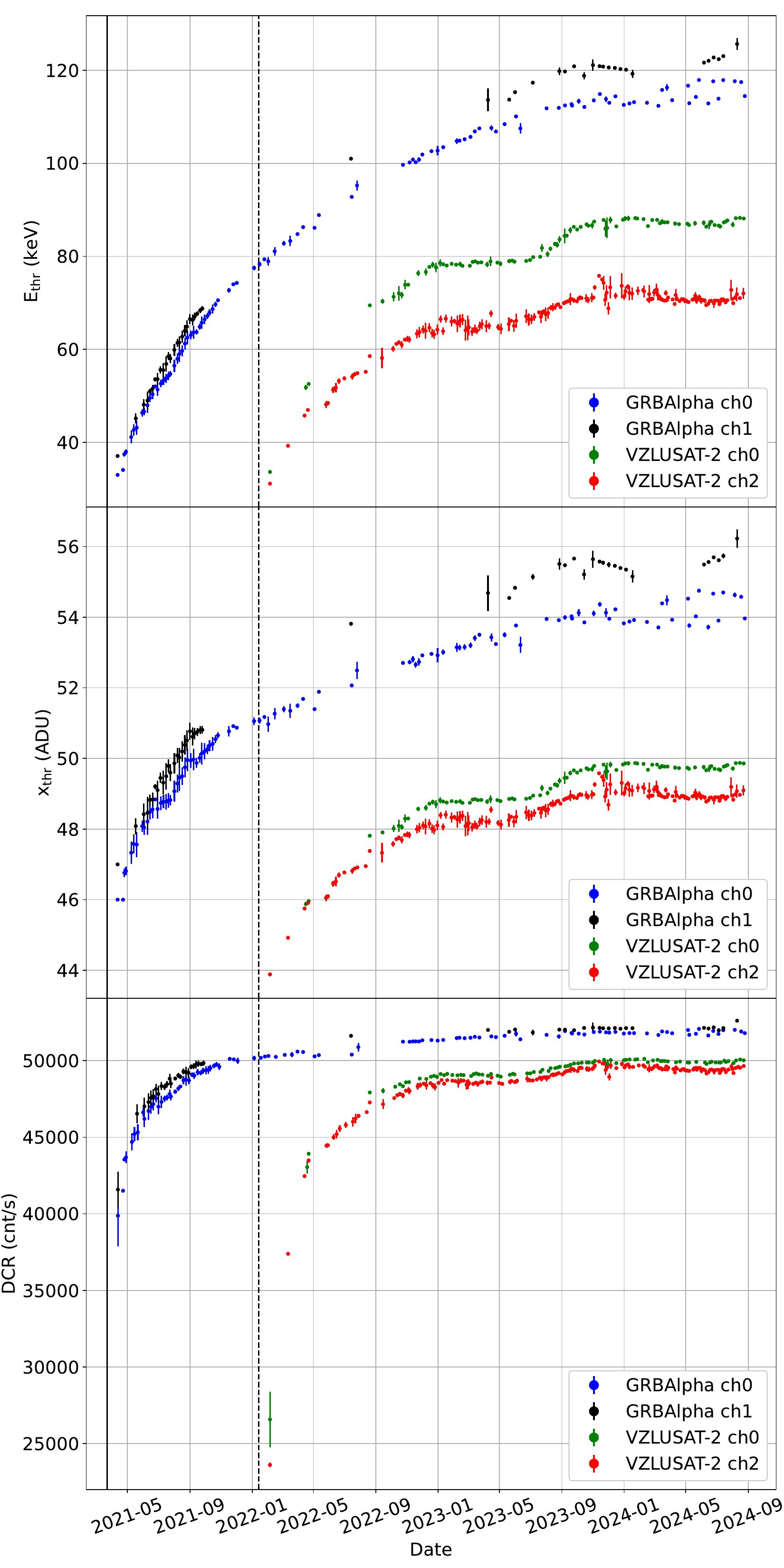}
	\caption{Top: Evolution of the low-energy threshold over the time on orbit for the gamma-ray detectors on the GRBAlpha and VZLUSAT-2 CubeSats. For GRBAlpha two readout channels \emph{ch0} and \emph{ch1} on one detector are shown. For VZLUSAT-2 a readout channel \emph{ch0} (\emph{ch2}) on detector unit 0 (1) is shown. Middle: Similar to the top panel, however here the threshold is plotted in terms of the directly measured spectral channel number (ADU). Bottom: Evolution of the dark count rate, i.e. the integrated spectrum over the noise peak. The vertical solid and dashed lines mark the launch date of GRBAlpha and VZLUSAT-2, respectively.}
	\label{fig:thr_evol}
\end{figure}

Fig.~\ref{fig:thr_evol} presents the evolution of the low-energy threshold over the time on orbit for the gamma-ray detectors on the GRBAlpha and VZLUSAT-2 CubeSats. The figure shows the evolution of the threshold in terms of directly measured spectral channel number (ADU), in terms of keV derived from the ADU channel number using the gain calibration described in Sec.~\ref{sec:methods_thr_dcr}, and the evolution of DCR. The derived threshold and DCR values obtained within five days were grouped together and the plot shows the mean values with the uncertainties calculated as corrected sample standard deviations of measured threshold or DCR within those five days.

It is necessary to say that the dark count rate reported in the datasheet of the MPPC S13360-3050 PE is $5\times10^5$\,cnt/s. For four MPPCs connected in parallel, the rate would be $2\times10^6$\,cnt/s. However, our measurement gives much lower values of $\sim2.5-5\times10^4$\,cnt/s. The reason is that the shaping parameter of the analogue electronics is $\approx10\,\mu$s. During the signal digitisation, the pulse processing is done within the resolution time of $15\,\mu$s \cite{2023A&A...677L...2R}. Therefore, the observed spectrum close to the tip of the noise peak is strongly affected by the pulse processing of the detector’s electronics resulting in a strong pulse pile-up and the actual rate of thermally-generated avalanches in all the MPPCs’ pixels is much higher than the measured dark count rate.

\begin{figure}[h!]
	\centering
        \includegraphics[width=0.99\linewidth]{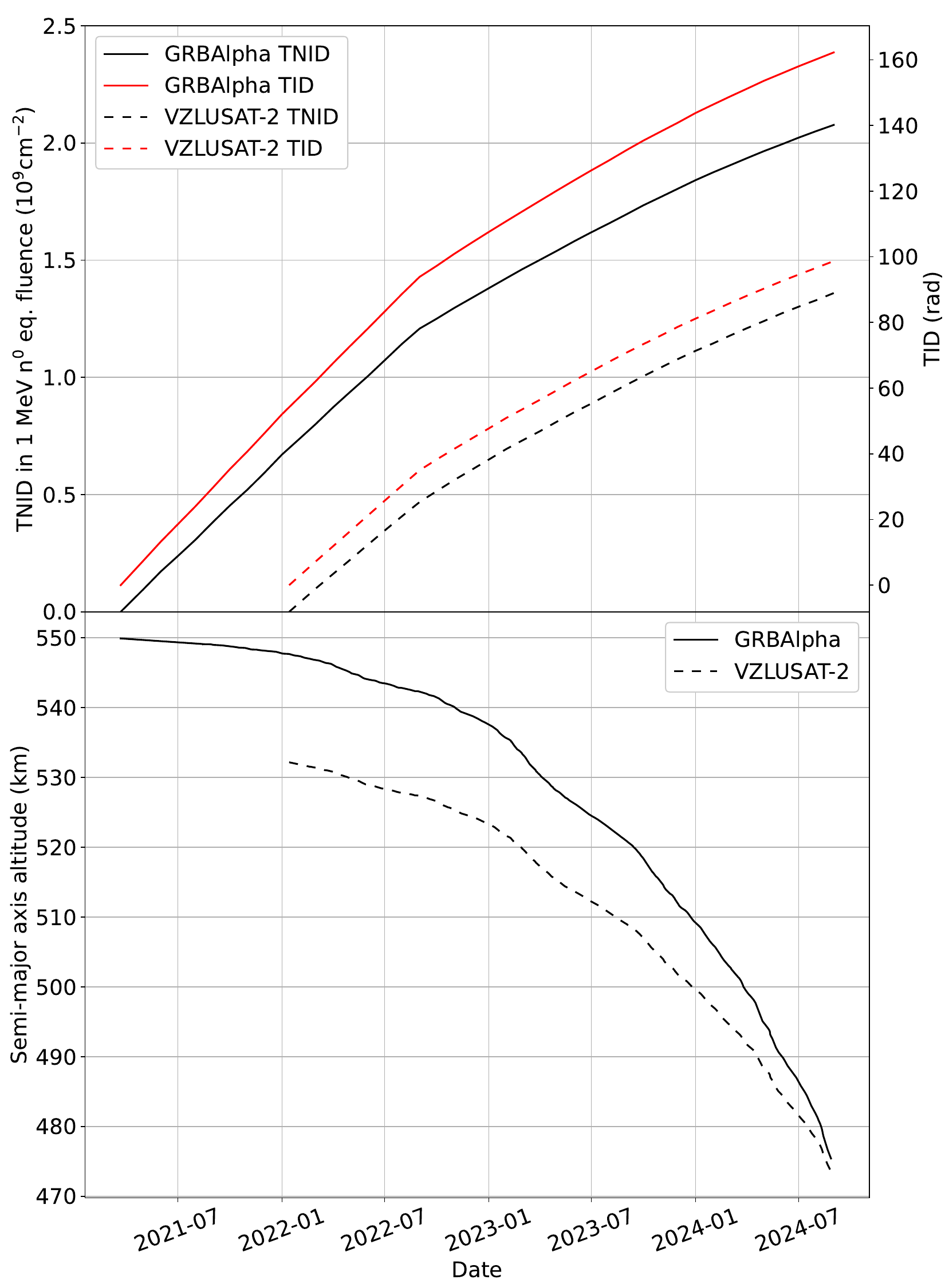}
	\caption{Top: Total ionizing dose (TID) and total non-ionizing dose (TNID) in Si simulated by the GRAS software using the AP-8 model of geomagnetically trapped protons and the actual orbital parameters of the CubeSats and their development is displayed. TNID is expressed in 1 MeV neutron equivalent fluence.} Bottom: A progress of the semi-major axis altitude for GRBAlpha and VZLUSAT-2.
	\label{fig:doses_alt_evol}
\end{figure}

Fig.~\ref{fig:doses_alt_evol} presents the result of the expected simulated TID and TNID deposited in Si for the gamma-ray detectors on the GRBAlpha and VZLUSAT-2 together with the progress of the semi-major axis altitude of both satellites.

The radiation damage in silicon is higher due to protons than due to electrons \citep{2014JInst...9P7016D, 2019NIMPA.926...69G, 2023NIMPA104567488A} on a single particle level, however it depends on shielding and spectrum. At LEO entering the inner Van Allen radiation belt, i.e. passing SAA might pose a higher risk of faster ageing of SiPMs. Our simulations confirm that this is true at LEO for the orbit of GRBAlpha and the $2-2.5$\,mm thick lead shielding protecting MPPCs in its GRB detector. At the beginning of the mission when the satellite was at the altitude of $\sim550$\,km and for the AP-8 and AE-8 models for the solar cycle minimum, the ratio of TID and TNID due to trapped protons and electrons was $\mathrm{TID(p^+)/TID(e^-) = 74\pm14}$ and $\mathrm{TNID(p^+)/TNID(e^-) = 2\,680\pm850}$, respectively. Later, when the satellite altitude dropped to ${\sim475}$\,km and when we used the AP-8 and AE-8 models for the solar cycle maximum the ratio was $\mathrm{TID(p^+)/TID(e^-)} = 16.1\pm3.4$ and $\mathrm{TNID(p^+)/TNID(e^-)} = 570\pm150$, respectively.

Also, the total ionizing dose from solar protons and Galactic cosmic rays is negligible compared to the dose due to the geomagnetically trapped protons \citep{2019AN....340..666R}, therefore in Fig.~\ref{fig:doses_alt_evol} we show only the TID and TNID due to the trapped protons.

\begin{figure}[h!]
    \centering
    \includegraphics[width=0.99\linewidth]{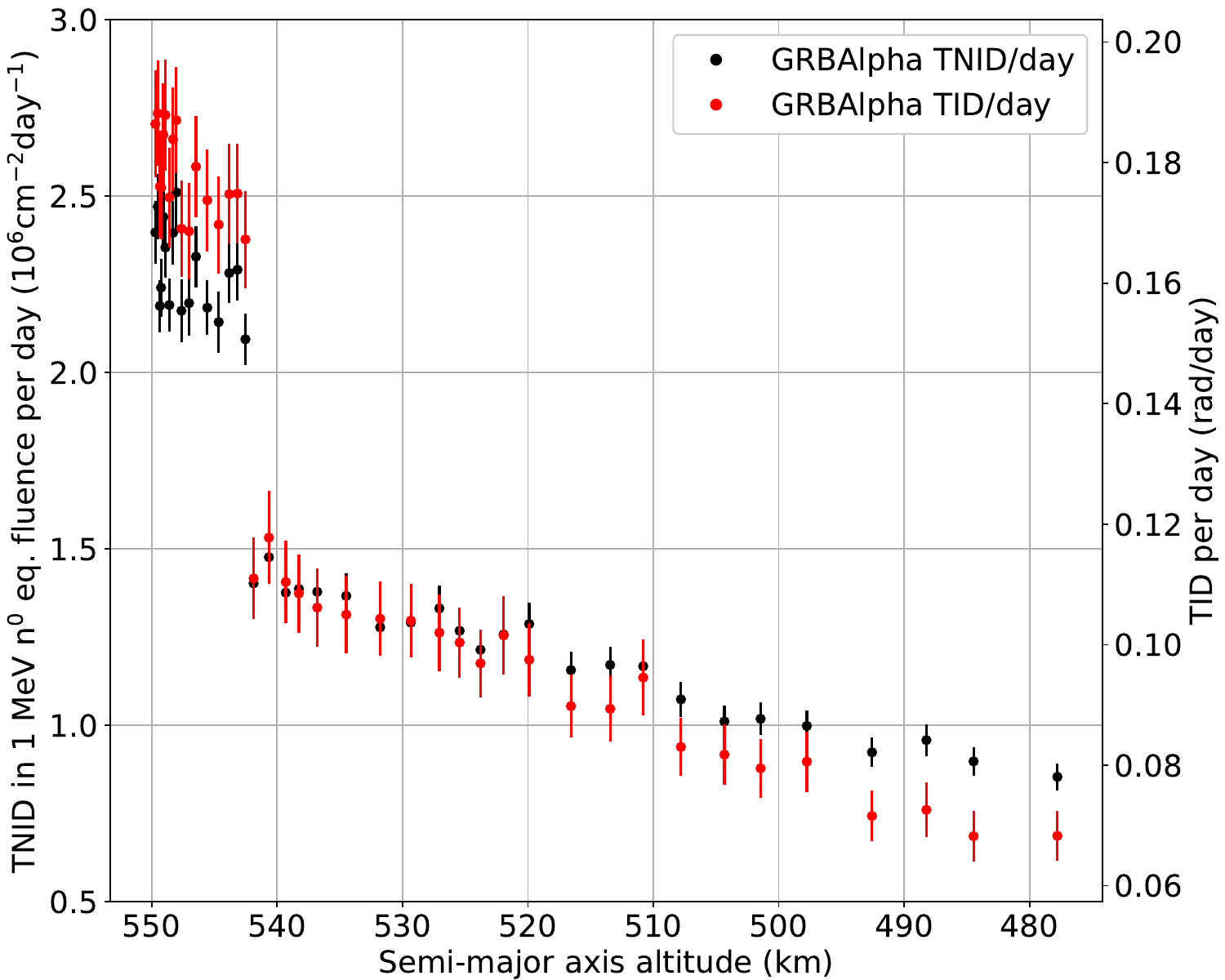}
    \caption{Total ionizing dose (TID) per day and total non-ionizing dose (TNID) in 1 MeV neutron equivalent fluence per day in Si simulated by the GRAS software using the AP-8 model of geomagnetically trapped protons and the actual orbital parameters of GRBAlpha. The drop at around 540\,km corresponds to the time when we changed the solar cycle activity from minimum to maximum in the simulation.}
    \label{fig:daily_doses_alt}
\end{figure}

At SSO and altitude of 550\,km the orbit-averaged integral flux of protons is about 60\,\% higher than at altitude of 500\,km according to AP-8 model (solar cycle minimum) of geomagnetically trapped protons \citep{2020SPIE11444E..3PR}. It is also known that there is a lower flux of trapped protons during the solar cycle maximum compared to the solar cycle minimum \citep{1996GMS....97..119H}. Fig.~\ref{fig:daily_doses_alt} presents the effect of lower altitude on TID and TNID deposited per day in shielded Si as obtained from simulations described in Sec.~\ref{sec:methods_expected_tid_nid} including a transition from the solar cycle minimum to maximum.
For example, from these simulations we obtained the dose in Si of an MPPC sensor in GRBAlpha's detector at the beginning of the mission during the first 21 days. That is between the launch and the measured background spectrum which became the basis of the template spectrum described in Sec.~\ref{sec:methods_thr_dcr}. The simulated TID was $3.9\pm0.2$\,rad and $\mathrm{TNID}=(5.0\pm0.2)\times10^7$\,cm$^{-2}$ 1\,MeV neutron equivalent fluence, respectively, which is about 2\,\% of the total dose in orbit since launch to the end of the period studied in this work (Sep. 2024). Similarly, for VZLUSAT-2 the simulations revealed for the first 22 days in orbit $\mathrm{TID}=4.0\pm0.2$\,rad and $\mathrm{TNID}=(5.4\pm0.2)\times10^7$\,cm$^{-2}$ 1\,MeV neutron equivalent fluence, i.e. about 4\,\% of the total dose until Sep. 2024.

\subsection{Temperature Dependence}
\label{sec:res-temp}

\begin{figure}[h!]
	\centering
	\includegraphics[width=0.99\linewidth]{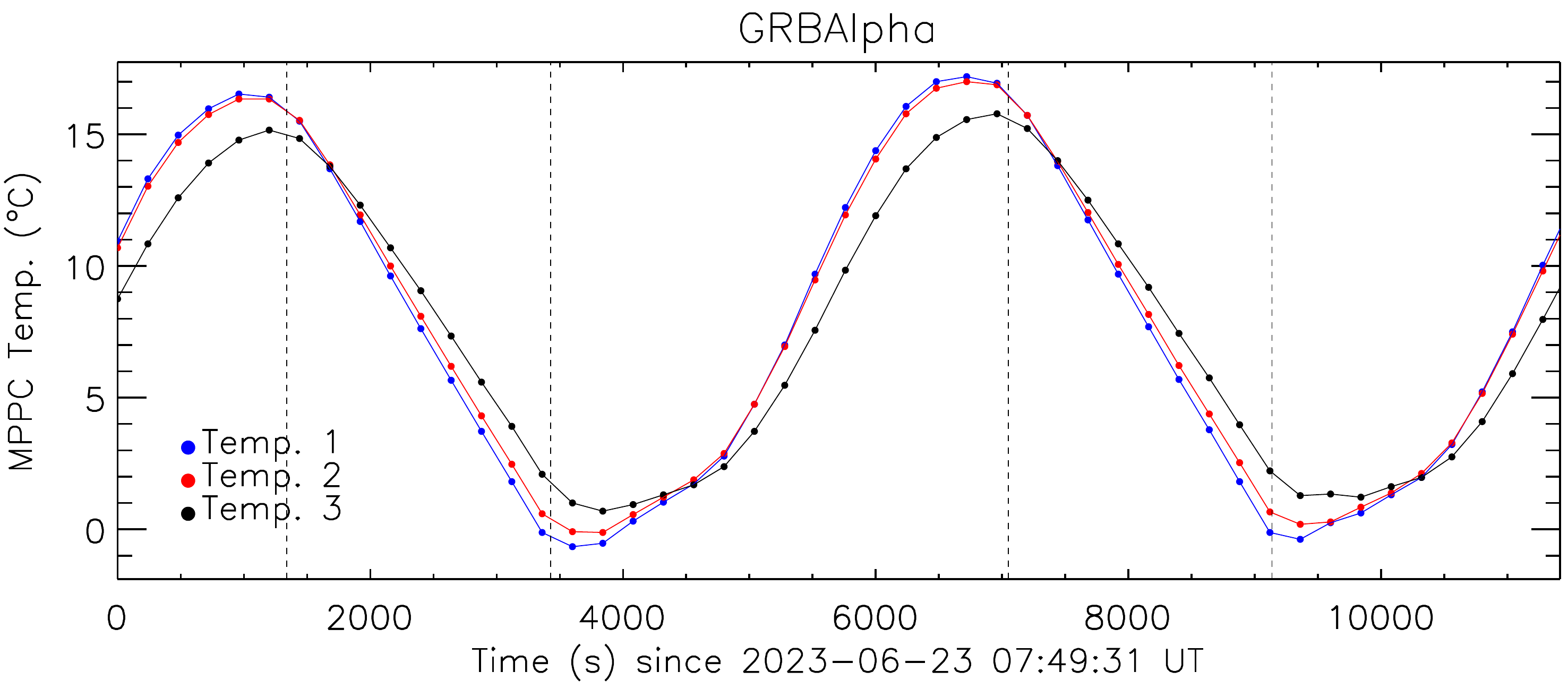}
        \includegraphics[width=0.99\linewidth]{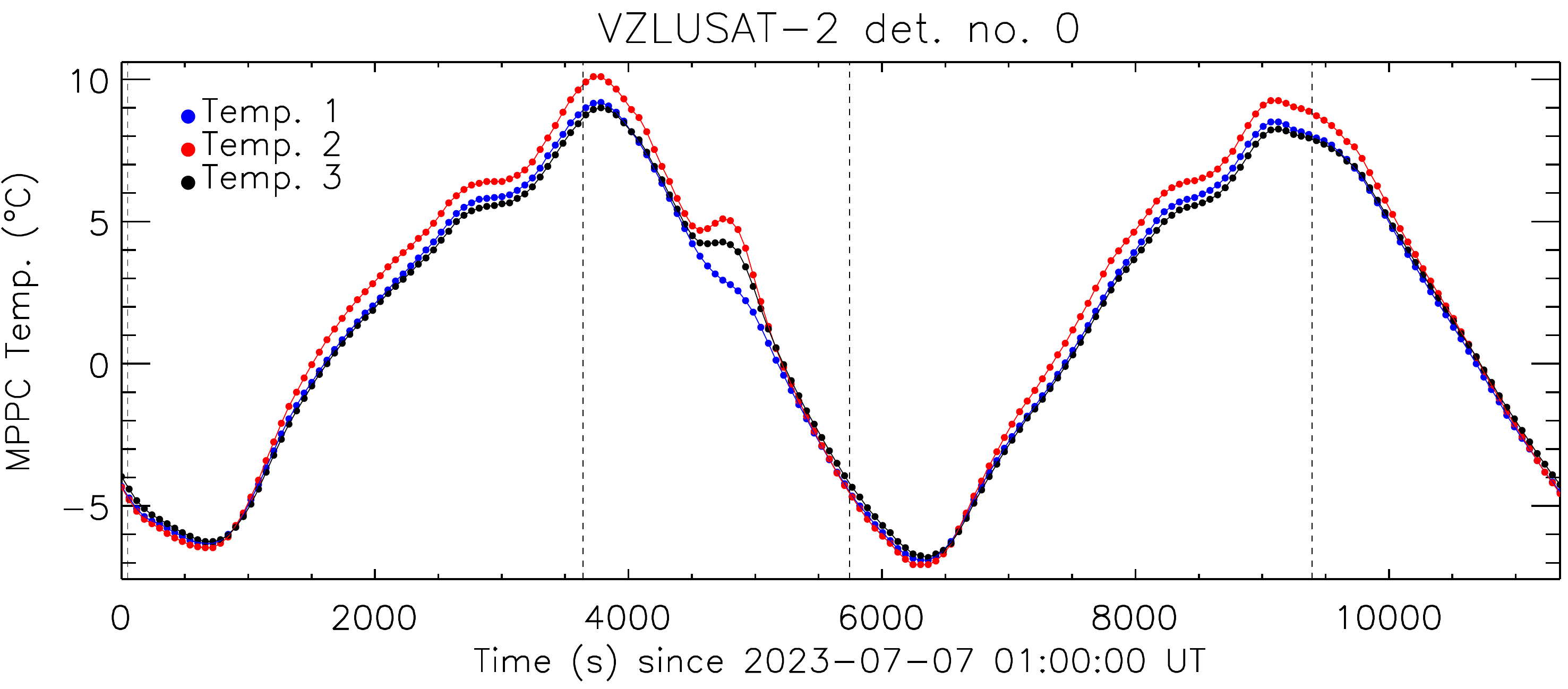}
        \includegraphics[width=0.99\linewidth]{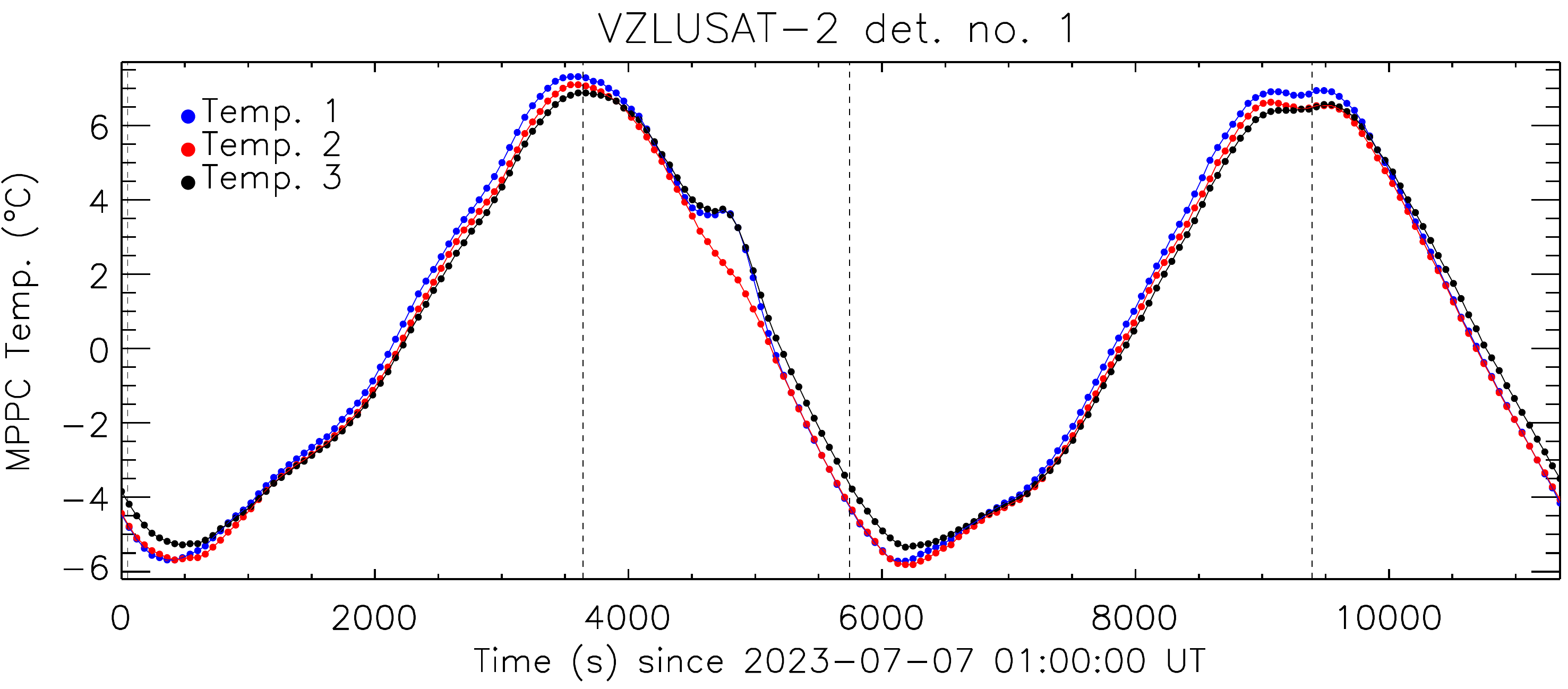}
	\caption{Measurements from three thermometers placed on the MPPC board of the GRBAlpha detector (Top), the VZLUSAT-2 GRB detector no. 0 (Middle), and the VZLUSAT-2 GRB detector no. 1 (Bottom), respectively. All measurements are shown over two orbits and the dashed lines mark the beginnings and ends of the solar eclipses according to the satellites' TLEs.}
	\label{fig:temp_grbalpha_vzlusat2}
\end{figure}

After the in-orbit upgrade of the payload software on GRBAlpha, we have simultaneous measurements of the temperature of the MPPC board together with the spectral measurements since Aug. 2022. The MPPC board is populated with three thermometers allowing us to measure a temperate gradient across the MPPC board. Therefore, we can study the dependence of the low-energy threshold and DCR on temperature in orbit. Fig.~\ref{fig:temp_grbalpha_vzlusat2} shows an example of the temperature variation of the MPPC boards on GRBAlpha and VZLUSAT-2 over two orbits.

\begin{figure}[h!]
	\centering
        \includegraphics[width=0.99\linewidth]{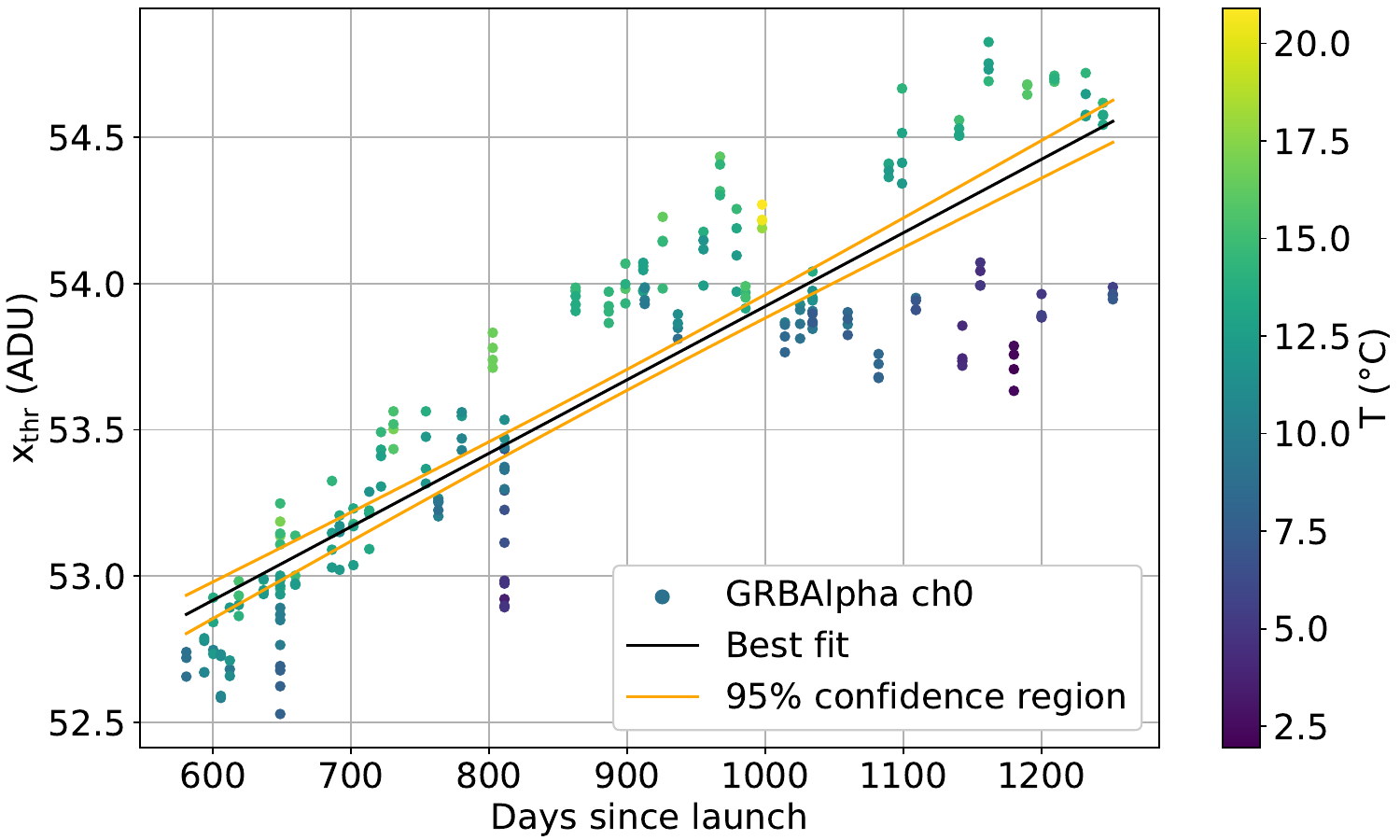}
	\caption{The noise peak threshold of the detector on GRBAlpha (readout channel \emph{ch0}) as a function of time in the interval where we have a simultaneous measurement of background spectra and the MPPC board temperature.}
	\label{fig:grbalpha_thr_time}
\end{figure}

\begin{figure}[h!]
	\centering
        \includegraphics[width=0.99\linewidth]{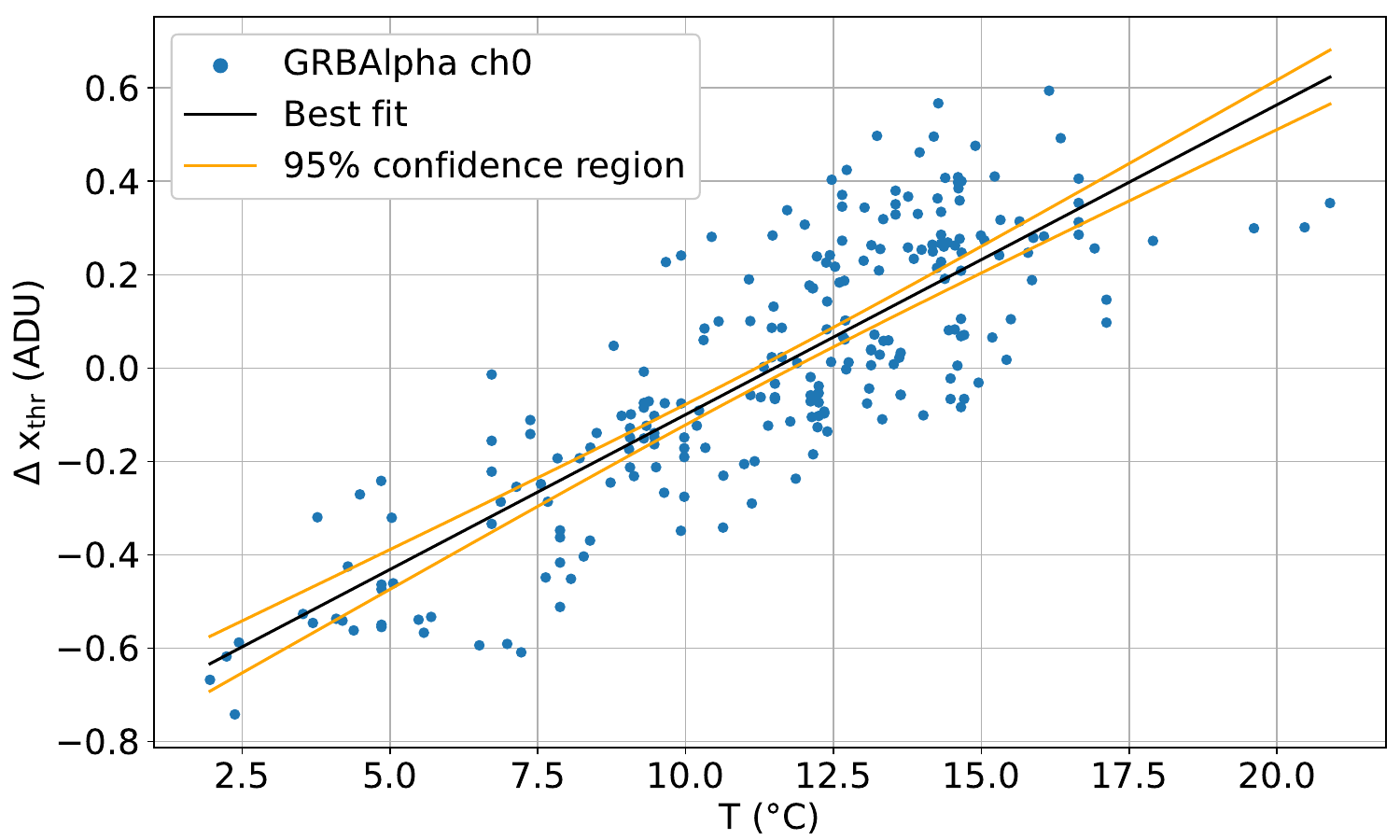}
	\caption{The change of the noise peak threshold of the detector on GRBAlpha (readout channel \emph{ch0}) with the average temperature of the MPPC board.}
	\label{fig:grbalpha_thr_res_temp}
\end{figure}

First, we fit the temporal dependence of the noise peak threshold $x_\mathrm{thr,fit}\mathrm{(ADU)}=A_0\,t+B_0$, where $t$ is the number of days in orbit since the launch of the satellite, see Fig.~\ref{fig:grbalpha_thr_time}. The best-fit parameters are $A_0=(2.51\pm0.09)\times10^{-3}\,\mathrm{ADU/d}$ and $B_0=(51.41\pm0.08)\,\mathrm{ADU}$. As seen from the figure a certain scatter of the low-energy threshold can be attributed to the temperature variation of the detector's MPPCs. Next, we subtract the temporal trend $\Delta x_\mathrm{thr}=x_\mathrm{thr}-x_\mathrm{thr,fit}$ and fit the residua with a linear function $\Delta x_\mathrm{thr}\mathrm{(ADU)}=A_1\,T+B_1$, where $T$ is the mean temperature in $^\circ$C over three thermometers of an MPPC board (see Fig.~\ref{fig:grbalpha_thr_res_temp}). The best-fit parameters are $A_1=(6.6\pm0.3)\times10^{-2}\,\mathrm{ADU}/^\circ\mathrm{C}$ and $B_1=(-0.76\pm0.04)\,\mathrm{ADU}$. The coefficient of determination is $R^2=0.68$, the Pearson correlation coefficient is $r=0.82$ and the $\mathrm{p-value}<10^{-10}$.

\begin{figure}[h!]
	\centering
        \includegraphics[width=0.99\linewidth]{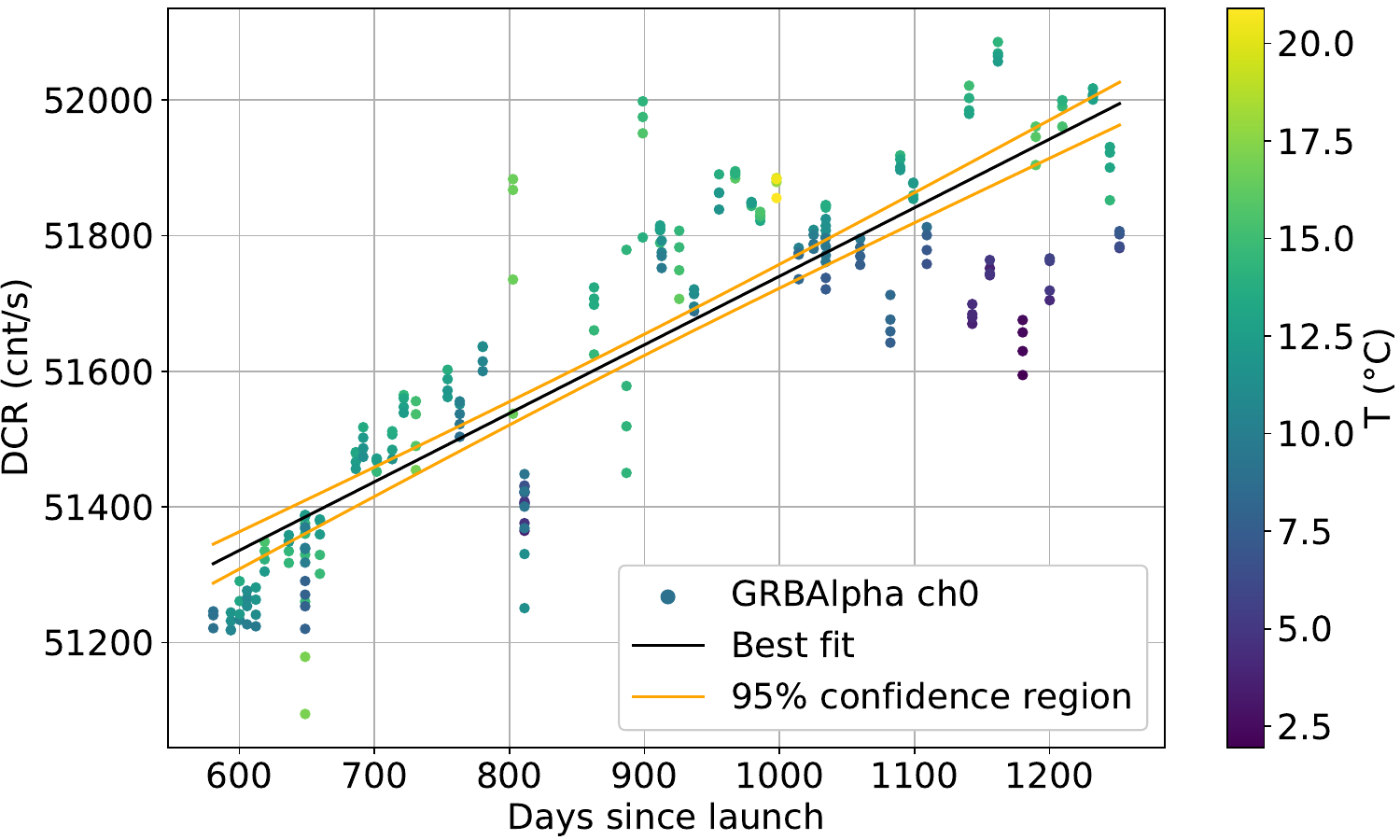}
	\caption{The dark count rate of the detector on GRBAlpha (readout channel \emph{ch0}) as a function of time in the interval where we have a simultaneous measurement of background spectra and the MPPC board temperature.}
	\label{fig:grbalpha_dcr_time}
\end{figure}

\begin{figure}[h!]
	\centering
        \includegraphics[width=0.99\linewidth]{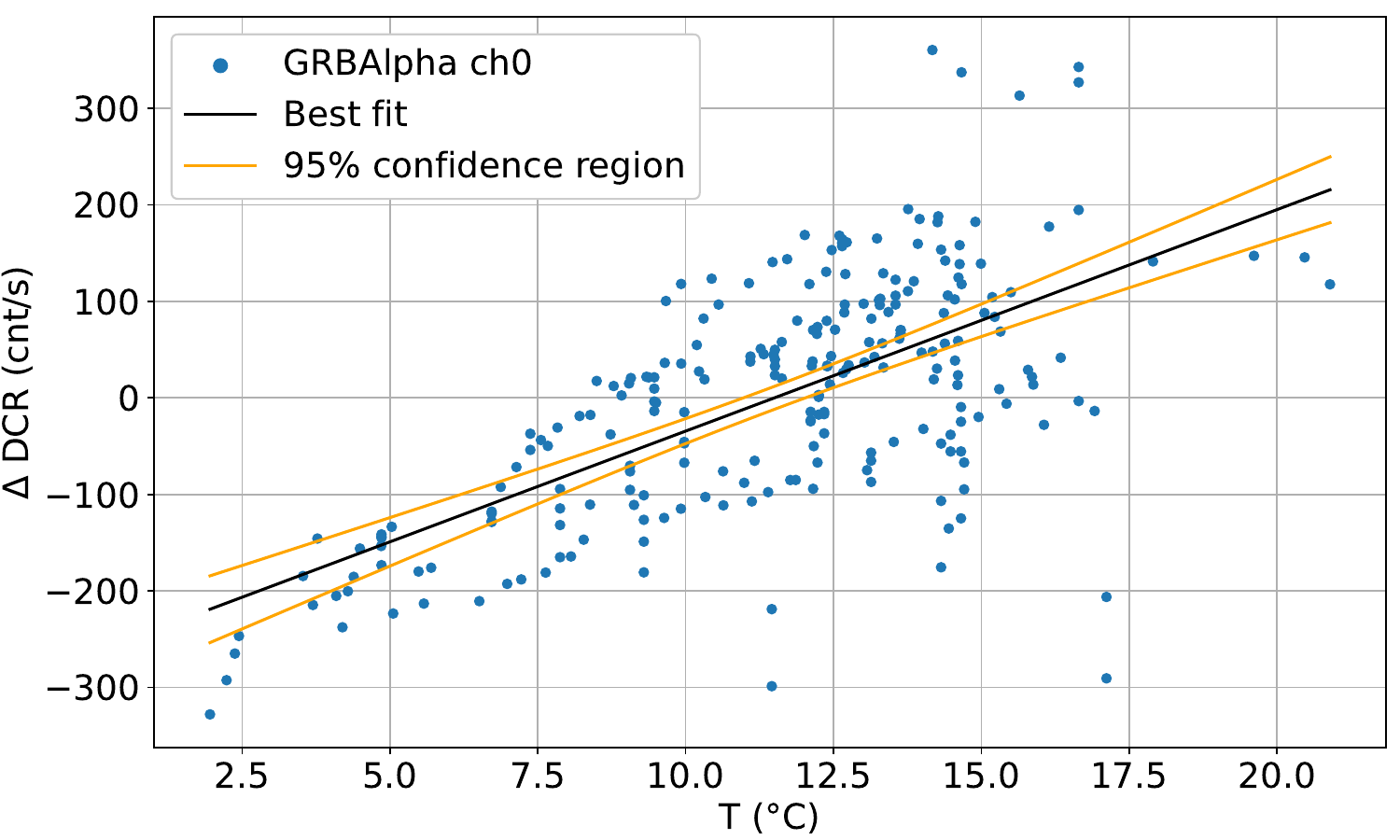}
	\caption{Change of the dark count rate of the detector on GRBAlpha (readout channel \emph{ch0}) with the average temperature of the MPPC board.}
	\label{fig:grbalpha_dcr_res_temp}
\end{figure}

Concerning DCR we proceed similarly. First, we fit the temporal dependence $DCR_\mathrm{fit}=C_0\,t+D_0$ (see Fig.~\ref{fig:grbalpha_dcr_time}). The best-fit parameters are $C_0=(1.01\pm0.04)\,\mathrm{cnt/s/d}$ and $D_0=(5.073\pm0.004)\times10^4\,\mathrm{cnt/s}$. Then we subtract the temporal trend $\Delta DCR=DCR-DCR_\mathrm{fit}$ and fit the residua with a linear function $\Delta DCR\mathrm{(cnt/s)}=C_1\,T+D_1$ (see Fig.~\ref{fig:grbalpha_dcr_res_temp}). The best-fit parameters are $C_1=(23\pm2)\,\mathrm{cnt/s}/^\circ\mathrm{C}$ and $D_1=(-260\pm20)\,\mathrm{cnt/s}$. The coefficient of determination is $R^2=0.42$, the Pearson correlation coefficient is $r=0.65$ and the $\mathrm{p-value}<10^{-10}$.

\section{Discussion}
\label{sec:discussion}

Notice that in Fig.~\ref{fig:noise_peak_spec_grbalpha} the VZLUSAT-2 detectors exhibit narrower noise peaks, which explains why the sensitivity thresholds and DCR of the VZLUSAT-2 detectors are lower, as shown in the middle and bottom panel of Fig.~\ref{fig:thr_evol}. This feature is also observed with the ground calibration, therefore it is likely by design due to different electronic gains or due to different scintillation light collection efficiency, for example, because of the use of optical glue vs optical bonding pads.

The change of the gain of the detector over the time the satellite spent at LEO as manifested in Fig~\ref{fig:grbalpha_gain_evolution} is due to increased dark current. In the electronics of the detectors on GRBAlpha and VZLUSAT-2 there is 1\,k$\Omega$ resistor present between the high voltage circuitry and the MPPCs to regulate the bias voltage, therefore increased dark current causes a drop of the bias voltage. Lower bias voltage means lower the gain of the detector. Therefore, the gain factor $g$ of the energy calibration shown in Fig~\ref{fig:grbalpha_gain_evolution} increases with time.

The observed plateauing of the time dependence of the sensitivity threshold and DCR shown in Fig.~\ref{fig:thr_evol} is interesting and we interpret that as it might be due to lower proton flux at lower altitudes as the orbital decay progresses.
Although GRBAlpha and VZLUSAT-2 do not have any technical system which would allow active annealing of MPPCs in their GRB detectors, to some extent, there might also be a small contribution due to long-term annealing occurring when the detectors are heated up by the solar radiation. We note that as explained in Sec.~\ref{sec:methods} we do not have enough measurements of the activation lines to track the evolution of the gain of detectors on VZLUSAT-2 and we have to rely only on the pre-launch calibration. Therefore, it is expected that the low energy threshold of VZLUSAT-2 detectors in terms of keV is in reality higher than what is displayed in the top panel of Fig.~\ref{fig:thr_evol}.

Also note that in Fig.~\ref{fig:thr_evol} in the case of GRBAlpha the measured DCR increased from 40\,kcps (21 days after launch) to 52\,kcps which is $\sim30$\,\% increase. In the case of VZLUSAT-2, the measured DCR increased from 24\,kcps (22 days since launch) to about double to 50\,kcps. GRBAlpha was launched at the semi-major axis altitude of 550\,km and VZLUSAT-2 was launched at the semi-major axis altitude of 535\,km. Fig.~\ref{fig:daily_doses_alt} shows how different the simulated TID and TNID per day for different altitudes are for the GRBAlpha's mass model. Even 15\,km altitude has a non-negligible difference in the exposed dose. It is possible that the DCR of GRBAlpha’s MPPCs was increasing faster after the launch. Unfortunately, we do not have available spectral measurements that could be used to explore the DCR immediately after launch due to the commissioning of various satellite subsystems.

In Fig.~\ref{fig:thr_evol} it can be seen that the low-energy threshold and DCR are higher for the readout channel $ch0$ compared to the readout channel $ch2$ of the VZLUSAT-2 detectors. We suspect that this difference is due to the higher total dark current of the four parallelly connected MPPCs of $ch0$ compared to $ch2$. The summed pre-irradiation dark current of the four MPPCs used for $ch0$ at a recommended bias voltage reported by the manufacturer in the delivery specifications sheet is 0.511\,$\mu$A, whereas for $ch2$ it is 0.482\,$\mu$A.

Both CubeSats at SSO transverse through SAA several times a day experiencing activation of the material in the detector due to irradiation by high-energy protons. During the duration of the GRBAlpha mission, we regularly collected high-resolution spectra after the satellite passed SAA northwards into the low-background region. This measurement was done typically every half year. With GRBAlpha we observe one short-term activation line at $260$\,keV for the nominal operating voltage of MPPCs. For lower operating voltage of MPPCs we observe, besides the $260$\,keV also the 511\,keV electron-positron annihilation line. Both lines vanish within a few minutes after the SAA passage.
The first observation of the activation line at $260$\,keV was done one month after the launch of GRBAlpha, specifically on 2021/04/20, when the degradation of MPPCs was small and the low-energy threshold was $\sim35$\,keV. Until the last measurement on 2024/10/22, we observed in the collected spectra only these two lines and we never observed any other line below $260$\,keV. Perhaps because the energy resolution on GRBAlpha is only about 30\,\%.
Moreover, the count rate of dark noise of MPPCs is much higher than the one from the activation of CsI or surrounding materials. Therefore, the additional count rate due to activation is negligible and it does not significantly affect our results of the low-energy threshold study.

The background spectra collected for this study were measured in low-background regions about minutes to $\sim20$\,min after the satellite passed the radiation belts with high concentrations of charged particles. Also, the datasheet of the CsI(Tl) manufacturer (Amcrys) states that the afterglow is 0.1\,\% after 6\,ms. Therefore, the afterglow in the scintillator should have only a marginal effect on our measurement results.

Concerning the energy calibration, one may ask why we do not study the single photoelectron energy spectrum of the collected measurements. There are two reasons. Events detected on board of both CubeSats are binned into 256 channels. The 256-bin spectrum is not sufficient to distinguish individual peaks of the single photoelectron spectrum. Moreover, after the degradation of MPPCs, the leakage current becomes much larger. The single photoelectron spectrum is not clear with most of the peaks being below the threshold.

\subsection{Comparison with Satellite's Housekeeping Data and Space Weather Activity}
\label{sec:res-weather}

As one can see from Fig.~\ref{fig:thr_evol} there is a period around Sep. 2023 when the low-energy threshold and DCR measured by read-out channels from both GRB detector units on VZLUSAT-2 manifest a faster increase which was preceded and followed by a rather constant level.

Therefore, we examined the housekeeping (HK) data of VZLUSAT-2 and searched for a relation. We examined HK data of the electrical power system (EPS): battery voltage, battery temperature, solar current and system current. Other HK data: the number of in-orbit on-board computer (OBC) resets and on-board radio PA temperature. We also checked the HK data concerning the communication with the satellite such as the number of received beacon frames and the number of received packets. However, we did not find any relation.

We also checked if there is any relation to the space weather activity using the Integrated Space Weather Analysis (ISWA) system provided by the Community Coordinated Modeling Center (CCMC)\footnote{\url{https://ccmc.gsfc.nasa.gov/tools/ISWA/}}. We compared: the index of magnetic activity derived from a network of near-equatorial geomagnetic observatories (DST), sunspot number, KP index; measurements performed by the Advanced Composition Explorer (ACE)\footnote{\url{https://science.nasa.gov/mission/ace/}} launched to L1 point which includes proton fluxes, electron fluxes, magnetic field, and solar wind bulk speed; measurements performed by the satellite GOES-P (formerly GOES-15)\footnote{\url{https://science.nasa.gov/mission/goes/}} including proton fluxes, electron fluxes, solar x-ray flux and magnetic field.
Furthermore, we inspected the electron and proton flux measurements made by the NOAA-15 satellite of the Polar Orbiting Environmental Satellites (POES)\footnote{\url{https://www.ospo.noaa.gov/Operations/POES/index.html}} program which is at SSO in the altitude of $\sim810$\,km and by the MetOp-B and MetOp-C\footnote{\url{https://www.eumetsat.int/our-satellites/metop-series}} satellites at SSO in the altitude of $\sim820$\,km. However, we did not find any coincidence between these space weather data and the shape of the progress of the low-energy threshold degradation measured by VZLUSAT-2.
Note that if this rapid increase of the low-energy threshold and DCR in measurements from VZLUSAT-2 around Sep. 2023 happened due to increased solar or space weather activity, then we expect it to be observed also in the data from GRBAlpha. However, it is not evident in the GRBAlpha’s measurements.

\section{Conclusions}
\label{sec:conclusions}

The SiMPs are prone to radiation damage which causes an increase in dark count rate and leads to an increase in the low-energy threshold of gamma-ray detectors. The increasing popularity of SiPMs among new spaceborne missions makes it important to characterize their long-term performance in the space environment. We have flight-proven SiMPs by Hamamatsu, MPPCs S13360-3050 PE, and we have studied their ageing due to radiation at LEO using measurements spanning over three years, which is, to the best of our knowledge, surpassing other missions. In particular, we have presented the evolution of the low-energy sensitivity threshold and dark count rate over time by using the in-orbit measurements by the GRBAlpha and VZLUSAT-2 CubeSats. For GRBAlpha the pre-launch low-energy threshold at $\sim20$\,keV degraded in 3 years and 5 months at LEO to $\sim120$\,keV which according to simulations by the GRAS tool corresponds to the accumulated TID of 160\,rad and TNID of 1 MeV neutron equivalent fluence of $2.1\times10^9$\,cm$^{-2}$ in Si due to geomagnetically trapped protons in Van Allen radiation belts (AP-8 model). We have demonstrated that MPPCs can be used in the LEO environment on a scientific mission lasting beyond three years when sufficient shielding of MPPCs is provided (in our case $\sim 2-2.5$\,mm thick PbSb alloy). This work manifests a potential of SiPMs being used in future high-energy astrophysics space missions.

\section*{Acknowledgements}
\label{sec:ack}
This work was supported by GAČR grant no. 24-11487J. We acknowledge the Community Coordinated Modeling Center (CCMC) at Goddard Space Flight Center for the use of the ISWA tool.





\bibliographystyle{elsarticle-num}
\bibliography{references.bib}







\end{document}